\documentclass[sigconf]{acmart}

\usepackage{bm}
\usepackage{algorithm}
\usepackage{algorithmic}

\AtBeginDocument{%
  }

\copyrightyear{2027}
\acmYear{2027}
\setcopyright{cc}
\setcctype{by}
\acmConference[WSDM '27]{Proceedings of the Twentieth ACM International Conference on Web Search and Data Mining}{February 15--19, 2027}{Hong Kong, Hong Kong}
\acmBooktitle{Proceedings of the Twentieth ACM International Conference on Web Search and Data Mining (WSDM '27), February 15--19, 2027, Hong Kong, Hong Kong}
\acmDOI{10.1145/3862322.3863822}
\acmISBN{979-8-4007-3117-4/2027/02}

\newcommand{\mcl}[1]{\mathcal{#1}}
\newcommand{\mbf}[1]{\mathbf{#1}}
\newcommand{\mbb}[1]{\mathbb{#1}}

\begin{document}
\settopmatter{printacmref=true}

\title{BACH: A Bayesian Admixture of Contrastive Heads for Multi-Interest Two-Tower Retrieval}

\author{Quoc Phong Nguyen}
\orcid{0000-0002-0183-2525}
\affiliation[obeypunctuation=true]{%
  \department{Amazon International Machine Learning}, \institution{Amazon}\\
  \country{Australia}}
\email{qphong@amazon.com}

\author{Paul Albert}
\orcid{0000-0001-6021-3962}
\affiliation[obeypunctuation=true]{%
  \department{Amazon International Machine Learning}, \institution{Amazon}\\
  \country{Australia}}
\email{albrtpa@amazon.com}

\author{Long Vuong}
\orcid{0000-0002-4745-2579}
\affiliation[obeypunctuation=true]{%
  \department{Amazon International Machine Learning}, \institution{Amazon}\\
  \country{Australia}}
\email{longvt@amazon.com}

\author{Vuong Le}
\orcid{0000-0003-1582-1269}
\affiliation[obeypunctuation=true]{%
  \department{Amazon International Machine Learning}, \institution{Amazon}\\
  \country{Australia}}
\email{levuong@amazon.com}

\author{Julien Monteil}
\orcid{0000-0002-8723-2677}
\affiliation[obeypunctuation=true]{%
  \department{Amazon International Machine Learning}, \institution{Amazon}\\
  \country{Australia}}
\email{jul@amazon.com}
\renewcommand{\shortauthors}{Nguyen et al.}

\begin{abstract}
Two-tower retrievers compress each user into one embedding, forcing unrelated
interests to share one representation. Multi-interest retrievers emit several
embeddings, but a common target-routed design has two limitations. First, target
routing sends direct contrastive gradient only through the nearest head and
cannot be reproduced at serving, where the best head scores each candidate.
Second, the unweighted maximum cannot express unequal user-specific mass across
heads. We introduce \emph{Bayesian Admixture of Contrastive Heads}~(BACH),
which treats the heads as latent retrieval components and fits a per-user
mixture of softmaxes with a variational lower bound. Soft
responsibilities send weighted contrastive gradients to every head rather than
one selected head, and a separate gating tower predicts the user-specific weights
a maximum cannot express. Amortized power-spherical or von Mises-Fisher
densities supply those responsibilities and per-head concentrations. The
corresponding spherical mixture reuses the same heads, gate, and concentrations
at serving, so neither stage uses target-selected hard routing. In sequential
full-catalog experiments on MovieLens-20M, Taobao, and Netflix, a controlled
ablation tests the routing mismatch: scoring every training candidate by its own
best head, as serving does, improves recall and mean average precision at every
head count on all three benchmarks. Against this stronger hard-routing control,
which shares BACH's encoder and item tower, personalized BACH improves both
metrics at every head count. With 32 heads and item embeddings fitted without
held-out interactions, the averages over five training seeds are also higher
for both metrics. It also matches or improves the displayed metrics
of single-vector models and of multi-interest baselines whose interest
extractors differ.
\end{abstract}



\maketitle

\section{Introduction}
\label{sec:introduction}

Two-tower models map user context and items into a shared embedding space for
large-catalog recommendation \cite{covington2016deep,yi2019sampling}. Dual
encoders use the same factorization for dense retrieval
\cite{karpukhin2020dense}. Because the item encoder is independent of user
context, item vectors can be precomputed and indexed for
\emph{approximate nearest-neighbor}~(ANN) search
\cite{yi2019sampling,karpukhin2020dense}, leaving
one user-vector computation on the online path. This design forces unrelated
interests into one embedding.

Multi-interest retrievers relax this bottleneck by producing $k$ user vectors
\cite{weston2013nonlinear,li2019multi,cen2020controllable,pal2020pinnersage}.
We use \emph{multi-interest} in this conventional architectural sense of
representing a user with several vectors. Without semantic labels or a
disentanglement objective, these vectors need not correspond to
human-interpretable topics.

A common design routes an observed target to its closest head and
uses that head to score every training candidate
\cite{cen2020controllable}. At serving, the best head instead scores each
candidate, and the system merges results by an unweighted maximum
\cite{weston2013nonlinear,cen2020controllable}. Section~\ref{sec:background}
formalizes these two rules.
First, target-selected routing sends an example's direct contrastive gradient
only through the selected head; the other heads receive no direct update from
that example. It also differs from the per-candidate score used at serving: a
negative near another head can be weak during training but strong in the merged
ranking.
Second, the unweighted maximum gives
every head equal influence and cannot express that different heads should carry
different user-specific mass.
For example, a head fitted to one atypical event can promote an item as
strongly as a head supported by many events.

We introduce \emph{Bayesian Admixture of Contrastive Heads}~(BACH), which
formulates multi-interest retrieval as a latent-variable mixture
(Section~\ref{sec:method}). Each head defines a contrastive softmax component,
and a separate gating tower predicts per-user mixture weights. Amortized
power-spherical recognition densities
\cite{decao2020power} or von Mises--Fisher recognition densities
\cite{davidson2018hyperspherical} provide soft responsibilities and per-head
concentrations. We optimize all parameters jointly through the
\emph{evidence lower bound}~(ELBO) \cite{blei2017variational}. The responsibilities send
weighted contrastive gradients to every head rather than only one selected head, while
the gate supplies the user-specific mass that an unweighted maximum cannot
express.
We use \emph{admixture} in the sense of a user-specific composition $\pi_u$
over latent heads: different heads can receive responsibility for different
interactions from the same user, rather than assigning the user as a whole to
one component. Here \emph{Bayesian} denotes latent-head inference under the
gate prior, not Bayesian parameter estimation.

The variational analysis in Section~\ref{sec:variational} connects this soft
router to contrastive training. For fixed generative parameters, maximizing the
ELBO over the recognition model minimizes its \emph{Kullback--Leibler}~(KL)
divergence from the exact contrastive posterior, while joint training also
adapts the contrastive components. Any remaining KL gap is a variational
approximation gap, not the target-selected routing mismatch above.

Section~\ref{sec:serving} carries this latent routing form into retrieval through
a normalized spherical mixture that reuses the same heads, gate, and
concentrations without a target-selected head. It also separates component
\emph{direction} from prior \emph{mass} to define personalized heads and a
shared global codebook under the same objective. Ordinary per-head ANN search
can generate candidates for the personalized model. For global BACH, head
candidate lists are precomputable while the gate remains personalized.

We evaluate BACH with exact full-catalog scoring on MovieLens-20M
\cite{harper2015movielens}, Taobao \cite{zhu2018learning}, and Netflix
\cite{bennett2007netflix} with shared preprocessing.
Section~\ref{sec:results} compares the complete BACH mixture with single-vector
and hard-routing models. A controlled ablation in
Section~\ref{sec:routing-analysis} holds the architecture and serving maximum
fixed while replacing target-only routing with per-candidate competition. This
change improves recall and \emph{mean average precision}~(MAP) at every tested
head count on all three datasets in the reported runs. Personalized BACH reports
an additional margin over this stronger hard-routing baseline.
The same analysis examines the learned gate mass and the breadth of recognition
responsibilities. We also evaluate the global-codebook form under exact
full-catalog scoring as a separate capacity configuration.

We make three contributions spanning the training objective, serving
construction, and empirical evaluation:
\begin{itemize}
  \item We formulate a variational mixture-of-softmaxes training objective with
  amortized power-spherical or von Mises--Fisher recognition. Soft
  responsibilities distribute weighted contrastive updates across heads, while
  a separate gate models user-specific component mass
  (Section~\ref{sec:method}).
  \item We derive a normalized spherical-mixture serving score that reuses the
  trained heads, gate, and concentrations without target-selected hard routing.
  Separating component direction from prior mass yields personalized and
  global-codebook forms; the latter makes head candidate lists precomputable
  while retaining personalized mixture weights
  (Section~\ref{sec:serving}).
  \item Under the full-catalog protocol on three sequential benchmarks,
  per-candidate hard routing improves R@100 and MAP over target-only routing at
  every tested head count. Personalized BACH improves both metrics over this
  control at every head count, has higher R@100 and MAP averaged over five
  seeds at $k=32$ under training-only item geometry, and matches or improves
  the displayed metrics of methods with different encoders or extractors.
  Learned gates assign nonuniform mass, while held-out-positive
  responsibilities remain distributed across multiple heads
  (Section~\ref{sec:experiments}).
\end{itemize}

\begin{figure*}[t]
\centering
\includegraphics[width=0.98\textwidth]{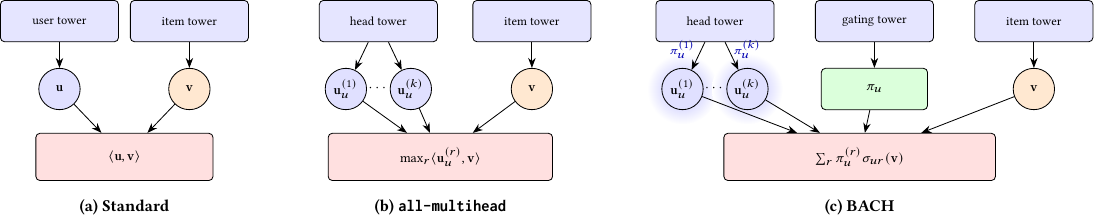}
\caption{Three retrieval formulations. A standard retriever uses one user
vector. The \texttt{all-multihead} control uses $k$ vectors and an unweighted
maximum; each candidate sends gradient only through its maximizing head. BACH
treats the heads as soft spherical components (blue halos) and predicts their
user-specific weights with a separate gate. Training uses soft responsibilities
over contrastive components; serving uses the displayed normalized spherical
mixture.}
\Description{A three-panel comparison of a single-vector two-tower model, an
unweighted hard multihead model, and BACH. The BACH heads have soft density
halos and user-specific mixture weights supplied by a separate gating tower.}
\label{fig:overview}
\end{figure*}

\section{Multi-Interest Two-Tower Retrieval}
\label{sec:background}

This section fixes notation for single-vector retrieval and then states the two
hard-routing objectives that BACH replaces and that serve as its closest
baselines.

\subsection{Single-Vector Retrieval}

Let $\mcl{U}$ and $\mcl{I}$ be users and items. A standard two-tower model
maps user context $\mbf{x}_u$ and item features $\mbf{y}_i$ to a common
$d$-dimensional space,
\begin{equation}
\mbf{u}_u=f_\theta(\mbf{x}_u),\qquad
\mbf{v}_i=g_\phi(\mbf{y}_i),\qquad
s(u,i)=\langle\mbf{u}_u,\mbf{v}_i\rangle.
\label{eq:two-tower}
\end{equation}
We $\ell_2$-normalize both vectors, so $s$ is a cosine similarity. Given observed
interactions $\mcl{D}=\{(u_n,i_n)\}_{n=1}^N$, a full-catalog softmax would
model
\begin{equation}
p(i\mid u)=
\frac{\exp(s(u,i)/\tau)}
{\sum_{j\in\mcl{I}}\exp(s(u,j)/\tau)}
\label{eq:full-softmax}
\end{equation}
with temperature $\tau$. The catalog normalizer is usually too expensive
during training. For each example, we instead form an indexed candidate
multiset $\mcl{C}_n$ from the positive, in-batch positives
\cite{yi2019sampling}, and uniformly sampled items \cite{yang2020mixed}, and
replace $\mcl{I}$ by $\mcl{C}_n$ in Equation~\eqref{eq:full-softmax}. Sums over
$\mcl{C}_n$ count repeated item IDs separately. The item vectors are computed
offline and indexed, leaving only $f_\theta$ and ANN search on the online path.

\subsection{Multiple Heads and Hard Routing}

A multi-interest tower
$F_\theta(\mbf{x}_u)=(\mbf{u}_u^{(1)},\ldots,\mbf{u}_u^{(k)})$ replaces the
single user vector while the item tower remains unchanged. For interaction
$n$, we abbreviate $\mbf{u}_{u_n}^{(r)}$ as $\mbf{u}_n^{(r)}$. This increases
representational capacity: an item can be near any head rather than a
compromise direction. Systems typically issue one ANN query per head and
merge candidates under
$s_{\max}(u,i)=\max_r\langle\mbf{u}_u^{(r)},\mbf{v}_i\rangle$.

\paragraph{Hard target routing.}
Hard-routing systems such as ComiRec \cite{cen2020controllable} select the head
nearest the positive,
\[
r_n^\star=\arg\max_r\langle\mbf{u}_n^{(r)},\mbf{v}_{i_n}\rangle,
\]
and use this one head for every item occurrence in $\mcl{C}_n$:
\begin{equation}
\mcl{L}_{\mathrm{pos}}
=-\sum_n\log
\frac{\exp(\langle\mbf{u}_n^{(r_n^\star)},\mbf{v}_{i_n}\rangle/\tau)}
{\sum_{j\in\mcl{C}_n}
 \exp(\langle\mbf{u}_n^{(r_n^\star)},\mbf{v}_j\rangle/\tau)}.
\label{eq:pos-routing}
\end{equation}
ComiRec uses the best-matching head as an item's base relevance score at serving
\cite{cen2020controllable}. For example, suppose the positive selects head 1,
while a sampled negative has dot products 0.1 and 0.9 with heads 1 and 2.
Equation~\eqref{eq:pos-routing} uses only the 0.1 similarity, giving the
negative a logit of $0.1/\tau$, whereas serving assigns it
$\max(0.1,0.9)=0.9$. The training loss therefore does not directly penalize the
0.9 similarity to head 2 that determines this item's serving score.

The positive's direct update also reaches only the selected head.
Consequently, a head that is never selected by a target receives no direct
contrastive update under this objective. Max serving is unweighted as well, so
it cannot assign less user-specific mass to one head than another; every head
can place an item at the top of the merged ranking.

\paragraph{Per-candidate max routing.}
A serving-consistent hard baseline scores every candidate with its best head:
\begin{equation}
\mcl{L}_{\mathrm{all}}
=-\sum_n\log
\frac{\exp(s_{ni_n}/\tau)}
{\sum_{j\in\mcl{C}_n}\exp(s_{nj}/\tau)},\quad
s_{nj}=\max_r\langle\mbf{u}_n^{(r)},\mbf{v}_j\rangle.
\label{eq:all-routing}
\end{equation}
We call Equations~\eqref{eq:pos-routing} and \eqref{eq:all-routing}
\texttt{pos-multihead} and \texttt{all-multihead}, and we refer to both as
hard routing. \texttt{all-multihead} routes candidate gradients through
different heads, but retains a non-smooth maximum and has no user-specific head
weights.

Comparing BACH only with \texttt{pos-multihead} would mix two changes:
replacing target routing with per-candidate competition, and replacing hard
competition with a variational mixture. We therefore retain both hard
objectives as separate baselines.

\section{BACH}
\label{sec:method}

BACH replaces hard routing with an explicit latent head variable. We first
define the contrastive generative model and its exact posterior, which depends
on the sampled candidate multiset. We then introduce a spherical recognition
family whose responsibilities do not depend on that multiset for fixed
parameters and are reusable at serving, derive the resulting lower bound, and
state the serving rule, the global-codebook variant, and the limiting cases.
Figure~\ref{fig:overview} provides a method overview by contrasting BACH's
gate-weighted spherical mixture with the single-vector retriever and
per-candidate hard routing.

\subsection{Contrastive Generative Model}

For each interaction, introduce a latent head
$z_n\in\{1,\ldots,k\}$ with
\begin{equation}
p(z_n=r\mid u_n)=\pi_{u_n}^{(r)},\qquad
\pi_u=\pi_{\bm{\xi}}(\mbf{x}_u)\in\Delta^{k-1},
\label{eq:prior}
\end{equation}
where $\Delta^{k-1}$ is the probability simplex over the $k$ heads and
$\pi_{\bm{\xi}}$ is a gating tower with parameters $\bm{\xi}$, separate from
$F_\theta$. Conditioned on $z_n=r$, head $r$ defines
\begin{equation}
p_r(i\mid u_n)=
\frac{\exp(\langle\mbf{u}_n^{(r)},\mbf{v}_i\rangle/\tau)}
{\sum_{j\in\mcl{C}_n}
 \exp(\langle\mbf{u}_n^{(r)},\mbf{v}_j\rangle/\tau)}.
\label{eq:component}
\end{equation}
The contrastive marginal is
$p(i_n\mid u_n)=\sum_r\pi_{u_n}^{(r)}p_r(i_n\mid u_n)$.
The observed-data objective is
\begin{equation}
\ell(\theta,\phi,\bm{\xi})
=\sum_{n=1}^N\log\sum_{r=1}^k
\pi_{u_n}^{(r)}p_r(i_n\mid u_n).
\label{eq:marginal}
\end{equation}

The softmax gate already enforces the simplex constraint on $\pi_u$, so
Equation~\eqref{eq:marginal} could be optimized directly. Its gradient weights
each head by the exact contrastive posterior below, and it gives the
concentrations used at serving no training signal. BACH instead chooses
a variational surrogate deliberately: it introduces a recognition family
whose responsibilities do not depend on the sampled candidate multiset and whose
component densities can also score the catalog at serving. The bound fits the
concentrations of that family through the variational gap
(Section~\ref{sec:variational}).

Without restricting responsibilities to the spherical recognition family, the
exact contrastive posterior responsibility is
\begin{equation}
\gamma_n^{\mathrm{ctr},(r)}
=\frac{\pi_{u_n}^{(r)}p_r(i_n\mid u_n)}
{\sum_q\pi_{u_n}^{(q)}p_q(i_n\mid u_n)}.
\label{eq:contrastive-posterior}
\end{equation}
The superscript $\mathrm{ctr}$ denotes the contrastive posterior. It includes a
different sampled-softmax denominator for every head. Its value changes when
$\mcl{C}_n$ changes, and reproducing the
corresponding mixture at serving would require $k$ full-catalog softmax
normalizers per request. The spherical recognition family below exchanges
bound tightness for routing that is independent of the catalog sampler.

\subsection{Spherical Recognition Model}

For normalized vectors, head $r$ is the mean direction of the spherical
component-density function
\begin{equation}
\sigma_{ur}(\mbf{v})=C_d(\kappa_{ur})
\omega_{\kappa_{ur}}\!
  \left(\langle\mbf{u}_u^{(r)},\mbf{v}\rangle\right),
\label{eq:spherical}
\end{equation}
where $\kappa_{ur}>0$ is the concentration. We study the power-spherical kernel
$\omega_\kappa(t)=(1+t)^\kappa$ \cite{decao2020power} and the
\emph{von Mises--Fisher}~(vMF) kernel
$\omega_\kappa(t)=\exp(\kappa t)$
\cite{banerjee2005clustering,davidson2018hyperspherical}. Both are monotone in
alignment. We retain the normalized density in both cases because
$C_d(\kappa)$ rescales components with different concentrations. The closed
forms and numerical evaluation are given in
Supplementary Section~\ref{app:derivations}.

\begin{figure}[t]
\centering
\includegraphics{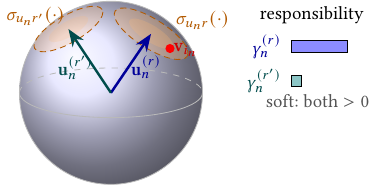}
\caption{Soft routing on the unit sphere (drawn for $d=3$). Each head is the
mean direction of a spherical density. Under the illustrated gate masses and
concentrations, the positive receives larger responsibility from the closer
head $r$, but both heads retain nonzero responsibility. Sampled negatives do
not enter this routing rule.}
\Description{A sphere with two head directions and their directional density
contours. A positive item lies nearer the first head. Bars show a large but
soft responsibility for that head and a smaller nonzero responsibility for
the other.}
\label{fig:spherical-routing}
\end{figure}

For interaction $n$, we amortize the concentration of head $r$ from its own
direction,
\begin{equation}
\kappa_{nr}:=\kappa_{u_n r}=\exp\!\left(
\rho_\psi(\operatorname{sg}[\mbf{u}_n^{(r)}])\right),
\label{eq:kappa}
\end{equation}
where $\rho_\psi$ is a network shared across heads and $\operatorname{sg}$ is
the stop-gradient operator, which prevents this path from updating the head.
The exponential keeps $\kappa_{nr}$ positive. The head still receives gradient
through the contrastive reconstruction and through its role as the spherical
mean direction in Equation~\eqref{eq:spherical}. Let
$\sigma_{nr}:=\sigma_{u_n r}(\mbf{v}_{i_n})$. The recognition responsibility is
\begin{equation}
\gamma_n^{(r)}
=\frac{\pi_{u_n}^{(r)}\sigma_{nr}}
{\sum_{q=1}^k\pi_{u_n}^{(q)}\sigma_{nq}}.
\label{eq:gamma}
\end{equation}
Figure~\ref{fig:spherical-routing} illustrates this routing rule. Each head
defines a directional density, and its gate mass modulates the resulting
responsibility. A positive can strongly favor one head while retaining nonzero
responsibility under the others.

Unlike the exact contrastive posterior, Equation~\eqref{eq:gamma} does not
reference sampled negatives. This makes routing stable for a fixed positive
and gives the same responsibility form as the serving mixture below. The learned
parameters can nevertheless remain sampler-dependent through the contrastive
likelihood in Equation~\eqref{eq:elbo}.

\begin{figure}[t]
\centering
\includegraphics{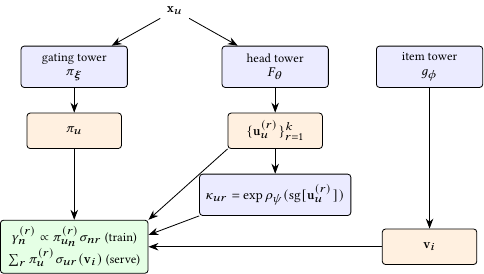}
\caption{BACH has separate head and gating towers. The concentration network
reads detached head directions. Soft spherical responsibilities route the
training ELBO; the corresponding spherical mixture scores serving items.}
\Description{A block diagram in which user history enters separate gating and
head towers. Head embeddings also enter a concentration network. Item
embeddings, mixture weights, heads, and concentrations define spherical
recognition responsibilities during training and the spherical mixture at
serving.}
\label{fig:architecture}
\end{figure}

Figure~\ref{fig:architecture} summarizes the separate computational paths:
history enters distinct head and gating towers, while each head direction also
determines its concentration. We predict the gate with a separate tower so that
prior mass is not tied to component geometry. Reusing a head-selection score as
the gate would make a broad, high-mass component difficult to distinguish from
a sharp, low-mass component. BACH uses separate parameters for direction
($F_\theta$), concentration ($\rho_\psi$), and prior mass
($\pi_{\bm{\xi}}$).

\subsection{Variational Objective}
\label{sec:variational}

For any categorical distribution $\gamma_n$ with full support, insert
$\gamma_n^{(r)}/\gamma_n^{(r)}$ into Equation~\eqref{eq:marginal} and apply
Jensen's inequality:
\begin{align}
\log\sum_r\pi_{u_n}^{(r)}p_r(i_n\mid u_n)
&=\log\sum_r\gamma_n^{(r)}
 \frac{\pi_{u_n}^{(r)}p_r(i_n\mid u_n)}
      {\gamma_n^{(r)}} \nonumber\\
&\geq \sum_r\gamma_n^{(r)}
 \log\frac{\pi_{u_n}^{(r)}p_r(i_n\mid u_n)}
          {\gamma_n^{(r)}}. \label{eq:jensen}
\end{align}
Substituting Equation~\eqref{eq:gamma} and summing examples yields
\begin{align}
\mcl{F}
=\sum_{n,r}\gamma_n^{(r)}
\bigl(\log\pi_{u_n}^{(r)}
&+\log p_r(i_n\mid u_n)
-\log\gamma_n^{(r)}\bigr). \label{eq:elbo}
\end{align}
The gap is
\begin{equation}
\ell-\mcl{F}
=\sum_n\operatorname{KL}
\left(\gamma_n\;\middle\|\;\gamma_n^{\mathrm{ctr}}\right),\quad
\gamma_n^{\mathrm{ctr},(r)}\propto
\pi_{u_n}^{(r)}p_r(i_n\mid u_n).
\label{eq:gap}
\end{equation}
Equation~\eqref{eq:gap} is the standard variational identity. For fixed
generative parameters, maximizing $\mcl{F}$ over the recognition model is
equivalent to minimizing this KL divergence from the exact contrastive
posterior in Equation~\eqref{eq:contrastive-posterior}. Joint optimization also
adapts the generative model toward a compatible contrastive marginal. Because
the recognition family is restricted and its optimization is amortized and
nonconvex, the bound need not be tight and we do not claim attainment of the
closest member of the family.

\paragraph{Objective interpretation.}
Suppressing interaction subscripts, the reconstruction term
$\sum_r\gamma_r\log p_r$ sends weighted contrastive
gradient to every head and the item tower. The prior term
$\sum_r\gamma_r\log\pi_r$ trains the gating tower. The entropy term
$-\sum_r\gamma_r\log\gamma_r$, together with the first two terms' dependence on
$\gamma$, trains the recognition model toward the contrastive posterior. Soft
responsibilities need not be large for every head, but unlike an argmax they
avoid structurally zero reconstruction weights.
Concentration is fitted through the variational gap, while the gate receives
gradient through both the prior and responsibilities.
Supplementary Section~\ref{app:derivations} further characterizes their
optimization roles.

\paragraph{Joint optimization.}
We maximize $\mcl{F}$ over $(\theta,\phi,\bm{\xi},\psi)$ with Adam
\cite{kingma2015adam} and differentiate through $\gamma$; responsibilities are
not detached targets. Since $z$ has only $k$ states, the expectation and entropy
are exact sums and require no Monte Carlo estimator or reparameterization.
Algorithm~\ref{alg:optimization} gives one minibatch update.

\begin{algorithm}[t]
\small
\caption{One minibatch BACH update. Gradients pass through the responsibilities
in line~\ref{line:elbo}; the method does not alternate detached E and M steps.}
\label{alg:optimization}
\begin{algorithmic}[1]
\REQUIRE batch $\{(u_n,\mbf{x}_{u_n},i_n)\}_{n=1}^B$, shared random-negative
multiset $\mcl{N}$, heads $k$, fixed $\tau$
\STATE Form the indexed multiset
$\mcl{C}=(i_1,\ldots,i_B,\mcl{N})$, retaining duplicate item IDs, and encode
its items.
\STATE Compute heads $\mbf{u}_n^{(r)}:=\mbf{u}_{u_n}^{(r)}$, gate
$\log\pi_n:=\log\pi_{u_n}$, and all
$B\times k\times|\mcl{C}|$ contrastive logits.
\STATE Gather $\log p_r(i_n\mid u_n)$ from each component softmax.
\STATE Compute $\kappa_{nr}$, $\log\sigma_{nr}$, and
$\log\gamma_n=\operatorname{logsoftmax}(\log\pi_n+\log\sigma_n)$, where the
$r$th entries of $\log\sigma_n$ and $\log\gamma_n$ are
$\log\sigma_{nr}$ and $\log\gamma_n^{(r)}$.
\STATE \label{line:elbo} Minimize
$-\frac{1}{B}\sum_{n,r}\gamma_n^{(r)}
[\log\pi_n^{(r)}+\log p_r(i_n\mid u_n)-\log\gamma_n^{(r)}]$.
\STATE Update $(\theta,\phi,\bm{\xi},\psi)$ jointly by Adam.
\end{algorithmic}
\end{algorithm}

\paragraph{Computational cost.}
Let $S=|\mcl{C}|$ count candidate positions, including repeated item IDs. BACH
and \texttt{all-multihead} both materialize $BkS$
scores, requiring $O(BkSd)$ dot-product work and $O(BkS)$ memory for the score
tensor. BACH adds $O(Bk)$ spherical and simplex operations and a gating tower,
but no sampling over $z$ or full-catalog normalizer. \texttt{pos-multihead}
evaluates only the selected head after routing, at the cost of the mismatch in
Equation~\eqref{eq:pos-routing}.

\subsection{Serving and Global Codebook}
\label{sec:serving}

At serving, BACH ranks items by the spherical density
\begin{align}
m(\mbf{v}_i\mid u)
&=\sum_{r=1}^k \pi_u^{(r)}\sigma_{ur}(\mbf{v}_i) \nonumber\\
&=\sum_{r=1}^k \pi_u^{(r)} C_d(\kappa_{ur})
\omega_{\kappa_{ur}}
\left(\langle\mbf{u}_u^{(r)},\mbf{v}_i\rangle\right).
\label{eq:serving}
\end{align}
For personalized heads, we retrieve the top $M$ items by inner product for
each head, union the candidates, and rerank the union by
Equation~\eqref{eq:serving}. Each component is monotone in its inner product,
so standard ANN indices can generate its candidates.

This continuous density on the unit sphere induces component responsibilities
of the form in Equation~\eqref{eq:gamma}, which Equation~\eqref{eq:gap} fits
toward the exact contrastive posterior during training and
Equation~\eqref{eq:serving} reuses as its exact component posterior at serving.
It is a ranking score for catalog items, not a discrete probability distribution
over the catalog or the sampled mixture of softmaxes in
Equation~\eqref{eq:marginal}.

BACH shares its heads, gate, and spherical component form across training and
serving, while normalizing components differently. The bound fits that
recognition family, including its concentrations, toward the exact contrastive
posterior; we do not claim an accuracy gain over direct optimization of
Equation~\eqref{eq:marginal}. At serving, the spherical components give a
closed-form, per-head ANN-compatible approximation to the catalog-normalized
mixture without its $k$ catalog normalizers per request.

This union reranking is exact only conditional on the retrieved candidates: an
item supported moderately by several components can be absent from every
component's top-$M$ list. Our main evaluation scores the full catalog exactly;
Section~\ref{sec:routing-analysis} measures this pipeline.

\begin{table}[t]
\centering
\caption{Online candidate-generation cost. $M$ candidates are requested per
ANN query; global lists and component scores are cached offline.}
\label{tab:serving-cost}
\small
\setlength{\tabcolsep}{2.1pt}
\begin{tabular}{lccc}
\toprule
Retriever & ANN vectors & ANN calls & Reranker \\
\midrule
Standard & 1 & 1 & inner product \\
Hard multihead & $k$ & $k$ & maximum \\
Personalized BACH & $k$ & $k$ & spherical mixture \\
Global BACH & 0 & 0 & cached scores \\
\bottomrule
\end{tabular}
\end{table}

As Table~\ref{tab:serving-cost} shows, personalized BACH issues the same $k$
ANN queries as a conventional $k$-head retriever. Its candidate union
$\mcl{A}$ contains at most $kM$ items, and evaluating all mixture terms costs
$O(k|\mcl{A}|d)$ if their dot products are recomputed. Global BACH instead
reweights cached lists and component scores. Larger $M$ increases each
component's candidate coverage at higher latency.

The objective also permits $k$ global codebook heads $\mbf{c}^{(r)}$, each with
$\lVert\mbf{c}^{(r)}\rVert_2=1$. They replace the outputs of
$F_\theta(\mbf{x}_u)$, so the heads and their ANN lists become
user-independent and can be computed offline;
online work reduces to predicting $\pi_u$ and reweighting cached lists. This
global-codebook form trades personalized directions for precomputation and a
fixed component codebook.
Supplementary Section~\ref{app:serving} defines a residual hybrid between these
forms.

\subsection{Properties and Limiting Cases}
\label{sec:properties}

\paragraph{Reduction to a single vector.}
When $k=1$, both $\pi_u$ and $\gamma_n$ equal one. The prior and entropy terms
in Equation~\eqref{eq:elbo} vanish, leaving the ordinary component
log-likelihood $\log p_1(i_n\mid u_n)$. At serving, either spherical kernel is
monotone in $\langle\mbf{u},\mbf{v}_i\rangle$ for fixed positive $\kappa$, so
it induces the same item ordering as a normalized single-vector two-tower
model. BACH therefore contains the standard retriever as a ranking and
training special case; its additional behavior begins only when several
components compete.

\paragraph{A weighted smooth maximum.}
For vMF components, it is convenient to rank in log space:
\begin{equation}
\log m(\mbf{v}_i\mid u)=
\operatorname{LSE}_{r}\!\left[
\log\pi_u^{(r)}+\log C_d(\kappa_{ur})
\quad+\kappa_{ur}\langle\mbf{u}_u^{(r)},\mbf{v}_i\rangle
\right],
\label{eq:smooth-max}
\end{equation}
where $\operatorname{LSE}$ is log-sum-exp. If all heads share a concentration
$\kappa$, then
$\kappa^{-1}\log m(\mbf{v}_i\mid u)$ approaches
$\max_r\langle\mbf{u}_u^{(r)},\mbf{v}_i\rangle-1$ as
$\kappa\rightarrow\infty$, because
$\kappa^{-1}\log C_d(\kappa)\rightarrow-1$. The additive constant is
item-independent, so the limiting ranking is the maximum-inner-product ranking.
Finite $\kappa$ instead lets several aligned heads contribute, while
$\log\pi_u^{(r)}$ shifts their relative mass. The power-spherical kernel has
the analogous log component
$\kappa\log(1+\langle\mbf{u}_u^{(r)},\mbf{v}_i\rangle)$. With learned
head-specific concentrations, BACH is not merely a temperature-smoothed max:
normalizers and slopes differ by component, allowing broad and sharp components
to coexist.
Supplementary Section~\ref{app:derivations} details exchangeability and stable
evaluation.

\section{Related Work}
\label{sec:related}

\paragraph{Multi-vector user representations.}
Sampled-softmax two-tower systems support large-catalog retrieval through
precomputed item vectors and ANN search
\cite{covington2016deep,yi2019sampling,yang2020mixed}. Multi-vector methods
increase user-side capacity but differ in both representation extraction and
scoring. MaxMF introduced multiple user factors
\cite{weston2013nonlinear}; \emph{Multi-Interest Network with Dynamic
Routing}~(MIND) extracts interests by capsule routing \cite{li2019multi}; and
ComiRec provides self-attentive and dynamic-routing
extractors \cite{cen2020controllable}. Octopus \cite{liu2020octopus} and
PinnerSage \cite{pal2020pinnersage} also build several representations from
behavior history. ColBERT applies a related late interaction to multi-vector
text retrieval \cite{khattab2020colbert}. Our proposed method instead changes
the training and serving objectives applied to the same user-side encoder as
the head-matched hard controls.

\paragraph{Target-conditioned training.}
MIND and ComiRec illustrate two target-conditioned objectives. MIND conditions
label-aware attention on the observed target \cite{li2019multi}. ComiRec hard
routing instead selects the interest nearest that target
\cite{cen2020controllable}. The target can guide either operation during
training but is unavailable before candidates are scored at serving.
The hard ComiRec form additionally sends the example's direct contrastive
update through only the selected head.

REMI studies a different failure mode
called routing collapse, in which each learned interest over-concentrates on
one or a few items from the behavior sequence, and regularizes the
item-to-interest routing weights \cite{xie2023rethinking}. MIRN uses an
EM-style sequence-to-interest routing step to construct interest capsules, but
its Capsule-aware training module then selects the capsule most compatible with
the label item \cite{zhang2023mirn}. BACH makes the interaction assignment a
latent variable and uses soft responsibilities, so every head receives a
responsibility-weighted contrastive update rather than only the target-selected
head.

\paragraph{Unweighted maximum merging.}
MaxMF defines user--item affinity by the maximum over a user's factors
\cite{weston2013nonlinear}, and ComiRec likewise uses the best-matching
interest as its retrieval score \cite{cen2020controllable}. This per-candidate
maximum does not itself create the target-routing mismatch: MaxMF already scores
each candidate with its maximizing factor, as does our \texttt{all-multihead} control.
It nevertheless supplies no user-specific weight for the factors, so an
otherwise low-mass head can dominate the ranking as readily as a high-mass one.
BACH replaces this unweighted merge with a gate-weighted spherical mixture.
The comparison with \texttt{all-multihead} separates that mixture from the
benefit of correcting target-only routing.

\paragraph{General probabilistic ingredients.}
The methods in this paragraph are general probabilistic modeling tools rather
than prior multi-interest retrieval systems. Mixtures of softmaxes increase the
expressiveness of conditional distributions in language modeling
\cite{yang2018breaking} and knowledge-graph link prediction
\cite{badreddine2026theoretical}. Variational inference provides a general
optimization framework for approximating a target posterior within a chosen
variational family \cite{blei2017variational}. BACH adapts these ingredients to
multi-interest retrieval and instantiates its components with normalized vMF
densities \cite{banerjee2005clustering,davidson2018hyperspherical} and
power-spherical densities \cite{decao2020power}. We use these densities to
define the serving score in Equation~\eqref{eq:serving}.

Diversity-aware reranking and subset-selection methods, such as maximal
marginal relevance \cite{carbonell1998use} and determinantal point processes
\cite{kulesza2012determinantal}, are complementary. BACH changes the relevance
score used to rerank a multihead candidate union but does not optimize an
explicit diversity objective.

\section{Experiments}
\label{sec:experiments}

\subsection{Protocol}

\paragraph{Experimental questions.}
The experiments address three questions. First, does the complete
variational mixture in Equations~\eqref{eq:gamma}, \eqref{eq:elbo},
and~\eqref{eq:serving} improve full-catalog retrieval over the compared
single-vector models and head-matched hard-routing controls? Second, how much
of the difference from
target-selected routing follows from replacing Equation~\eqref{eq:pos-routing}
with per-candidate competition in Equation~\eqref{eq:all-routing}? Third, do the
learned gate and recognition responsibilities exhibit the unequal component
mass and soft routing represented by $\pi_u$ and $\gamma_n$? The global-codebook
configuration documents a separate serving and capacity operating point from
Section~\ref{sec:serving}.

\paragraph{Data and preprocessing.}
We convert MovieLens-20M \cite{harper2015movielens}, Taobao UserBehavior
\cite{zhu2018learning}, and Netflix Prize \cite{bennett2007netflix} into
time-ordered sequences of unique items. We sort each user's interactions by
timestamp and deduplicate user--item pairs by retaining their first
occurrence. MovieLens-20M and Netflix retain ratings of at least four; Taobao
retains page views, carts, favorites, and purchases as implicit feedback. Before
splitting, we rank items by frequency over the complete retained data and cap
the catalog at 50,000, which preserves all qualifying MovieLens-20M and Netflix
items and truncates Taobao's long tail. We also keep users with at least 80
retained items and, where necessary, sample at most 100,000
users. Supplementary Table~\ref{tab:data} reports the resulting datasets.

\paragraph{Splits and evaluation.}
We use chronological interaction-level validation and test holdouts. For each
evaluated user, we score every eligible catalog item, with evaluation capped at
10,000 qualifying users. Supplementary Section~\ref{app:implementation}
details these splits and histories.

\paragraph{Models and training.}
All learned methods use normalized 64-dimensional vectors and the same frozen
PureSVD item initialization \cite{cremonesi2010performance} followed by a shared
\emph{multilayer perceptron}~(MLP) with a residual connection. Standard is a
single-vector baseline in which one learned summary token attends to all
observed history positions through a noncausal Transformer layer.
\emph{Self-Attentive Sequential Recommendation}~(SASRec)
\cite{kang2018self} is also single-vector, but it uses two causal self-attention
blocks and reads the final sequence position. The comparison therefore changes
the history encoder and its sequential inductive bias, not the number of
retrieval vectors.

Both hard controls and BACH use Standard's summary-token
encoder, while MIND and ComiRec retain their method-specific extractors.
Within ComiRec, one interest extractor uses \emph{dynamic routing}~(DR).
The other uses \emph{self-attention}~(SA). We label the corresponding variants
ComiRec-DR and ComiRec-SA. The nonlearned
\emph{most-popular baseline}~(MostPop) ranks items by training-set frequency
after excluding request-history items.
The item tower, candidate sampler, optimizer, and evaluation protocol are
matched across learned methods.

We sweep $k\in\{8,16,32\}$ for personalized multi-interest models and
$k\in\{128,256\}$ for global BACH; Supplementary
Section~\ref{app:implementation} gives the full configurations. We denote the
personalized power-spherical and vMF
variants by p-BACH and v-BACH, and the global power-spherical variant by global
p-BACH.

\paragraph{Shared preprocessing and item geometry.}
Catalog construction and user eligibility use complete retained sequences
before the chronological split. We fit the frozen PureSVD table once to the
complete retained interaction matrix, including the validation and test
interactions, and share it across every learned method.
On Taobao, the same retained data determine membership in the
frequency-truncated catalog. Thus the complete-matrix comparisons use the same
cohort, catalog, and item geometry. Their held-out exposure is shared across
learned methods, which controls relative comparisons but does not remove leakage
and may inflate absolute metrics relative to training-only item geometry.
Table~\ref{tab:train-only} repeats Standard and three $k=32$ multi-interest
configurations with training-only item geometry over five seeds.

\paragraph{Metrics and selection.}
We report MAP, Recall@$K$, and
\emph{normalized discounted cumulative gain}~(NDCG)@$K$
\cite{jarvelin2002cumulated}. We compute each metric
per user and report the mean across users. BACH ranks with
Equation~\eqref{eq:serving}; the other learned multihead methods rank by maximum
inner product. For every method, we order eligible items by decreasing score.
For user $u$, let $\mcl{G}_u$ be the ten test items, $\mcl{R}_{u,K}$ the first
$K$ items of this order, $\operatorname{rank}_u(i)$ one plus the number of
eligible items scored strictly above item $i$, and
$i_{u,(1)},\dots,i_{u,(10)}$ the test items in increasing order of this rank.
We compute per-user recall and \emph{average precision}~(AP) as
\begin{equation}
\operatorname{Recall@K}_u=
\frac{|\mcl{G}_u\cap\mcl{R}_{u,K}|}{|\mcl{G}_u|},\qquad
\operatorname{AP}_u=
\frac{1}{|\mcl{G}_u|}
\sum_{t=1}^{|\mcl{G}_u|}\frac{t}{\operatorname{rank}_u(i_{u,(t)})}.
\label{eq:metrics}
\end{equation}
Without score ties, $\operatorname{AP}_u$ is the mean precision at the positions
of the test items in the full eligible ranking. Exactly tied items share their
best rank in AP, while the order within a tied group is unspecified for
Recall@$K$ and NDCG@$K$. Identical frozen item vectors always produce such ties
for the learned methods. At most 85 of the 100,000 MovieLens-20M test targets,
2 on Netflix, and none on Taobao belong to an identical-vector group.
NDCG@$K$ uses binary
relevance and the ideal \emph{discounted cumulative gain}~(DCG) for ten
relevant items. Request-history items are excluded; validation and test targets
remain eligible.

Validation MAP selects checkpoints and $k$ within each model family.
Supplementary Table~\ref{tab:selected-k} reports the selected head counts;
standard and SASRec have no head-count sweep.
Tables~\ref{tab:head-matched} and~\ref{tab:routing} retain every tested $k$.
Test metrics are reserved for final reporting, not checkpoint or head-count
selection.

\subsection{Main Results}
\label{sec:results}

\begin{table*}[t]
\centering
\caption{Full-catalog test retrieval, best $k$ in each family by validation
MAP. ML20M is MovieLens-20M. R10/R100 are Recall@10/100; N100 is NDCG@100.
All learned results use the shared preprocessing and PureSVD item geometry
described in Section~\ref{sec:experiments}. The best displayed
learned result in each dataset/metric is bold. Ties for the best displayed
value are also bold. Global p-BACH shares 128 (ML20M) or 256 (Taobao, Netflix)
heads across users, whereas the personalized multi-interest rows use at most 32
heads per user, so these rows are not capacity-matched.}
\label{tab:main}
\scriptsize
\setlength{\tabcolsep}{3.1pt}
\begin{tabular}{lcccccccccccc}
\toprule
& \multicolumn{4}{c}{ML20M}
& \multicolumn{4}{c}{Taobao}
& \multicolumn{4}{c}{Netflix} \\
\cmidrule(lr){2-5}\cmidrule(lr){6-9}\cmidrule(lr){10-13}
Method & R10 & R100 & N100 & MAP
       & R10 & R100 & N100 & MAP
       & R10 & R100 & N100 & MAP \\
\midrule
MostPop       & .017 & .132 & .064 & .020 & .004 & .025 & .013 & .003 & .014 & .113 & .053 & .015 \\
MIND          & .023 & .164 & .082 & .025 & .009 & .063 & .031 & .008 & .019 & .128 & .065 & .019 \\
ComiRec-DR    & .031 & .196 & .101 & .031 & .011 & .073 & .037 & .010 & .049 & .234 & .130 & .042 \\
ComiRec-SA    & .045 & .253 & .134 & .042 & .019 & .104 & .056 & .015 & .066 & .285 & .165 & .056 \\
\texttt{pos-multihead} & .060 & .308 & .168 & .056 & .020 & .113 & .061 & .017 & .074 & .305 & .179 & .063 \\
Standard      & .066 & .331 & .182 & .062 & .025 & .127 & .070 & .020 & .088 & .329 & .200 & .075 \\
SASRec        & .068 & .338 & .187 & .065 & .025 & .126 & .070 & .020 & .095 & .347 & .213 & .081 \\
\texttt{all-multihead} & .070 & .339 & .188 & .066 & \textbf{.029} & .137 & .077 & \textbf{.023} & .095 & .339 & .210 & .080 \\
p-BACH        & \textbf{.075} & \textbf{.350} & \textbf{.196} & \textbf{.069}
              & \textbf{.029} & .141 & .079 & \textbf{.023}
              & .102 & .359 & .225 & .088 \\
v-BACH        & .073 & .346 & .193 & .068
              & \textbf{.029} & \textbf{.142} & \textbf{.080} & \textbf{.023}
              & \textbf{.106} & \textbf{.363} & \textbf{.229} & \textbf{.091} \\
global p-BACH & .053 & .297 & .159 & .053
              & .017 & .102 & .053 & .014
              & .083 & .318 & .194 & .072 \\
\bottomrule
\end{tabular}
\end{table*}

Table~\ref{tab:main} supports two comparison tiers. The hard-routing controls share
BACH's encoder, item tower, sampler, and protocol, so the comparison with
\texttt{all-multihead} isolates the change from hard routing to the complete
BACH mixture. MIND and ComiRec keep their own extractors, so their rows compare
overall performance rather than one component. A personalized BACH variant
matches or exceeds every displayed learned result on all three datasets. Against
\texttt{all-multihead}, p-BACH raises MovieLens-20M R@100 from .339 to .350 and
MAP from .066 to .069. On Taobao, v-BACH raises R@100 from .137 to .142, while
R@10 and MAP match at the displayed precision. Netflix has the largest displayed
separation: v-BACH reaches .363 R@100 and .091 MAP, compared with .339 and .080
for \texttt{all-multihead}.

At its 128/256-head shared-codebook setting, global p-BACH exceeds MIND and ComiRec-DR on every displayed metric across all three
datasets, and exceeds ComiRec-SA on MovieLens-20M and Netflix with slightly lower
Taobao values. Personalized p-BACH remains stronger, consistent with the
additional flexibility provided by user-conditioned head directions.

Computed from the three-decimal values reported here, the relative change over
\texttt{all-multihead} on MovieLens-20M is 7.1\% at R@10, 3.2\% at R@100, and
4.5\% in MAP. Netflix shows a larger 11.6\% R@10 and 13.8\% MAP increase, but
only 7.1\% at R@100. On Taobao, the displayed MAP values match after rounding,
although v-BACH improves R@100 and NDCG@100. The larger relative differences at
R@10 and MAP than at R@100 on Netflix are compatible with cross-head reranking,
but BACH jointly changes the gate, concentration, spherical merge, and
variational training. These results therefore compare BACH and hard routing as
complete methods.

At unrounded precision, the Recall@10--NDCG@100 Pareto set of every dataset
contains only personalized BACH variants (Supplementary
Figure~\ref{fig:pareto-metrics}).

MIND and ComiRec have lower metrics than summary-token Transformer models under
this geometry. The comparison between \texttt{pos-multihead} and
\texttt{all-multihead} removes that encoder difference and directly tests the
routing rule.

\begin{table}[t]
\centering
\caption{Top: head-matched p-BACH and \texttt{all-multihead} results
(R@100/MAP); bold marks the higher unrounded value in each pair. Bottom:
differences (p-BACH minus \texttt{all-multihead}) at $k=32$ with 95\% paired
user-bootstrap confidence intervals.}
\label{tab:head-matched}
\scriptsize
\setlength{\tabcolsep}{2pt}
\begin{tabular}{lcccccc}
\toprule
& \multicolumn{2}{c}{MovieLens-20M}
& \multicolumn{2}{c}{Taobao}
& \multicolumn{2}{c}{Netflix} \\
\cmidrule(lr){2-3}\cmidrule(lr){4-5}\cmidrule(lr){6-7}
$k$ & \texttt{all} & p-BACH & \texttt{all} & p-BACH
    & \texttt{all} & p-BACH \\
\midrule
8  & .338/.065 & \textbf{.347/.068}
   & .135/.022 & \textbf{.138/.023}
   & .338/.080 & \textbf{.348/.083} \\
16 & .339/.066 & \textbf{.346/.068}
   & .137/.023 & \textbf{.140/.023}
   & .334/.079 & \textbf{.355/.086} \\
32 & .341/.067 & \textbf{.350/.069}
   & .136/.022 & \textbf{.141/.023}
   & .339/.080 & \textbf{.359/.088} \\
\bottomrule
\end{tabular}

\smallskip
\begin{tabular}{lcc}
\toprule
Dataset & $\Delta$R@100 [95\% CI] & $\Delta$MAP [95\% CI] \\
\midrule
MovieLens-20M & +.0093 [.0071, .0115] & +.0024 [.0017, .0032] \\
Taobao        & +.0048 [.0033, .0064] & +.0010 [.0006, .0015] \\
Netflix       & +.0201 [.0177, .0227] & +.0073 [.0061, .0084] \\
\bottomrule
\end{tabular}
\end{table}

Table~\ref{tab:head-matched} compares p-BACH with \texttt{all-multihead} at
each head count, removing both head-count and spherical-family selection;
Supplementary Figure~\ref{fig:performance-vs-k} plots this sweep together with the
target-routing control. p-BACH has higher unrounded
R@100 and MAP in all nine pairs, although Taobao MAP at $k=16$ is equal when rounded.
On MovieLens-20M, p-BACH gains 3.1--4.5\% relative MAP across head counts. On
Netflix, its MAP advantage is larger at $k=16$ and $32$ than at $k=8$, reaching
.088 versus .080 at $k=32$. This pattern reduces the possibility that the main
result follows only from selecting a favorable $k$. The hard maximum does not
improve monotonically with $k$: Taobao peaks at $k=16$, while Netflix changes
little.

At $k=32$, we use a paired user bootstrap over the 10,000 evaluation users,
with 10,000 resamples, to estimate 95\% percentile confidence intervals for
these differences, shown in the bottom panel of Table~\ref{tab:head-matched}.
Every interval excludes zero, including Taobao, where the displayed MAP values
differ by .001. The gain is not uniform across users: p-BACH has higher
per-user AP for 53\%, 55\%, and 57\% of MovieLens-20M, Taobao, and Netflix
users, respectively. These intervals condition on one trained model per method;
Table~\ref{tab:train-only} reports variation over five training seeds under
training-only item geometry.

\begin{table}[t]
\centering
\caption{Training-only PureSVD: mean and sample standard deviation over five
training seeds for Standard and three $k=32$ multi-interest configurations,
with the same 10,000 evaluation users per dataset. Bold
marks the higher mean of p-BACH and \texttt{all-multihead}. Exact seed-level
tests are reported in the text.}
\label{tab:train-only}
\scriptsize
\setlength{\tabcolsep}{3pt}
\begin{tabular}{lccc}
\toprule
& MovieLens-20M & Taobao & Netflix \\
\midrule
\multicolumn{4}{l}{R@100} \\
Standard & .3257$_{\pm.0017}$ & .1078$_{\pm.0003}$ & .3247$_{\pm.0030}$ \\
\texttt{pos-multihead} & .2693$_{\pm.0035}$ & .0897$_{\pm.0010}$ & .2960$_{\pm.0063}$ \\
\texttt{all-multihead} & .3373$_{\pm.0022}$ & .1208$_{\pm.0007}$ & .3334$_{\pm.0010}$ \\
p-BACH & \textbf{.3462}$_{\pm.0023}$ & \textbf{.1224}$_{\pm.0010}$ & \textbf{.3568}$_{\pm.0032}$ \\
\midrule
\multicolumn{4}{l}{MAP} \\
Standard & .0612$_{\pm.0003}$ & .0159$_{\pm.0002}$ & .0740$_{\pm.0018}$ \\
\texttt{pos-multihead} & .0440$_{\pm.0009}$ & .0119$_{\pm.0001}$ & .0641$_{\pm.0018}$ \\
\texttt{all-multihead} & .0649$_{\pm.0005}$ & .0186$_{\pm.0002}$ & .0795$_{\pm.0012}$ \\
p-BACH & \textbf{.0683}$_{\pm.0006}$ & \textbf{.0190}$_{\pm.0002}$ & \textbf{.0876}$_{\pm.0015}$ \\
\bottomrule
\end{tabular}
\end{table}

\paragraph{Training-only geometry and seed variation.}
To remove held-out co-occurrence from the item geometry, we refit PureSVD
without the validation and test interactions of every user eligible for
evaluation and retrain Standard and the three $k=32$ multi-interest
configurations with five seeds (Table~\ref{tab:train-only}). p-BACH has higher mean R@100 and MAP than
\texttt{all-multihead} on every dataset. Every p-BACH seed exceeds every
\texttt{all-multihead} seed on both metrics for MovieLens-20M and Netflix
(unadjusted exact two-sided permutation test of the seed-level mean difference,
$p=.008$). On Taobao, the R@100 ranges overlap ($p=.024$), whereas the MAP
ranges do not ($p=.008$). In matched seed-0 runs, the
refit lowers Taobao R@100 by .017--.018 for p-BACH and \texttt{all-multihead}
and changes it by at most .0061 on MovieLens-20M and Netflix. Test targets
assigned zero PureSVD rows account for at most 0.18\%.
Supplementary Table~\ref{tab:train-only-main} reruns the selected
configurations of Table~\ref{tab:main} with seed 0.

\subsection{Routing and Scope Analyses}
\label{sec:routing-analysis}

\paragraph{Placement of the maximum.}
Table~\ref{tab:routing} holds the architecture fixed and changes only
Equations~\eqref{eq:pos-routing} and \eqref{eq:all-routing};
Supplementary Figure~\ref{fig:performance-vs-k} visualizes the same sweep. Per-candidate
routing reports higher R@100 and MAP at every $k$ on every dataset; under
training-only item geometry at $k=32$, every \texttt{all-multihead} seed also
exceeds every \texttt{pos-multihead} seed on both metrics (unadjusted exact
two-sided permutation $p=.008$ in all six comparisons;
Table~\ref{tab:train-only}). The gap
grows with $k$ on MovieLens-20M and Taobao. Per-candidate routing also uses the
same max-over-head score in training and serving.

The denominator explains the failure mechanism. Suppose the positive is
closest to head 1 while a sampled negative is closest to head 2. Under
target-only routing, both are scored by head 1, so the negative may contribute
little loss even though head 2 will promote it at serving.
Equation~\eqref{eq:all-routing} penalizes that negative by applying the maximum
separately to every candidate. It also routes gradient through each candidate's
maximizing head, rather than only the positive-selected head.

At $k=32$, this controlled routing gap exceeds the remaining BACH margin over
\texttt{all-multihead}. MAP increases from .047 to .067 on MovieLens-20M, .013
to .022 on Taobao, and .063 to .080 on Netflix. Consequently,
\texttt{pos-multihead} is not an adequate sole baseline for BACH; part of an
apparent soft-routing gain can be obtained simply by correcting hard routing.
Using \texttt{all-multihead} as the hard comparator removes this denominator
effect.

\paragraph{Implications for target-routed training.}
Target-routed training, as in ComiRec's target-selected head and MIND's
label-aware attention \cite{cen2020controllable,li2019multi}, scores the
training candidates through a user vector selected or weighted by the positive.
In this controlled setting, per-candidate routing has higher R@100 and MAP at
every $k$, and its advantage increases with $k$ on MovieLens-20M and Taobao.
This change adds no parameters and uses the serving-time maximum during
training. An efficient target-only
implementation can score one selected head, whereas per-candidate routing
materializes a $k$-times larger score tensor. Target-routed baselines can therefore understate
what their heads can retrieve, and new multi-interest objectives should be
compared with per-candidate routing rather than with target routing alone.

\noindent\begin{minipage}{\columnwidth}
\centering
\captionsetup{hypcap=false}
\captionof{table}{Controlled routing (R@100/MAP): \texttt{pos} selects by the
positive; \texttt{all} routes each candidate.}
\label{tab:routing}
\scriptsize
\setlength{\tabcolsep}{2pt}
\renewcommand{\arraystretch}{0.82}
\begin{tabular}{lcccccc}
\toprule
& \multicolumn{2}{c}{MovieLens-20M}
& \multicolumn{2}{c}{Taobao}
& \multicolumn{2}{c}{Netflix} \\
\cmidrule(lr){2-3}\cmidrule(lr){4-5}\cmidrule(lr){6-7}
$k$ & \texttt{pos} & \texttt{all}
    & \texttt{pos} & \texttt{all}
    & \texttt{pos} & \texttt{all} \\
\midrule
8  & .308/.056 & \textbf{.338/.065}
   & .113/.017 & \textbf{.135/.022}
   & .305/.063 & \textbf{.338/.080} \\
16 & .302/.052 & \textbf{.339/.066}
   & .108/.015 & \textbf{.137/.023}
   & .298/.060 & \textbf{.334/.079} \\
32 & .278/.047 & \textbf{.341/.067}
   & .098/.013 & \textbf{.136/.022}
   & .300/.063 & \textbf{.339/.080} \\
\bottomrule
\end{tabular}
\end{minipage}
\par

\paragraph{Exchanged serving rules.}
Rescoring the $k=32$ checkpoints without retraining
(Supplementary Table~\ref{tab:serving-rules}), p-BACH's spherical score is at
most .008 R@100 below the catalog-normalized softmax mixture (.359 versus .367
on Netflix), which requires $k$ catalog normalizers per request. The
largest-term, $\pi_u^{(r)}=1/k$ mixture, and unweighted-maximum rules reduce
p-BACH R@100 by .011--.019, .008--.056, and .030--.069, respectively; the latter
two fall below \texttt{all-multihead} on every dataset. For
\texttt{all-multihead}, the uniform soft maximum and catalog-normalized softmax
mixture both lower R@100 relative to the default unweighted maximum.

\paragraph{Diversity, candidate generation, and sensitivity.}
p-BACH's top-100 lists contain the most distinct items and have the highest
tail-item coverage on every dataset (Supplementary Table~\ref{tab:diversity}). With per-head HNSW retrieval of $M=100$ candidates
and union reranking, R@100 equals its full-catalog value at the displayed
precision, and MAP decreases by at most .007. p-BACH has higher R@100 and MAP
than \texttt{all-multihead} at all three $M$ values, with median single-thread
CPU latency at $M=100$ of 5.8--8.4 versus 3.5--4.7 ms (Supplementary
Table~\ref{tab:ann}). On Taobao, p-BACH has higher R@100 than
\texttt{all-multihead} for $\tau\in\{0.03,0.05,0.1\}$ and $d\in\{32,128\}$, but
not at $\tau=0.2$ (Supplementary Table~\ref{tab:sensitivity}).

\paragraph{Learned mixture behavior.}
For personalized p-BACH at $k=32$, the median effective head count of the gate,
$\exp(H(\pi_u))$ with Shannon entropy $H$, is 18.17, 28.21, and 17.10 on
MovieLens-20M, Taobao, and Netflix, below the uniform value of 32. Held-out
positives need a median of 13, 15, and 9 heads, respectively, to accumulate
90\% recognition responsibility, rather than the single head imposed by hard
target routing (Supplementary Figure~\ref{fig:mechanism}). These diagnostics do
not assign semantic identities to heads.

\paragraph{Spherical family and concentration.}
Table~\ref{tab:main} reports higher displayed metrics for p-BACH on
MovieLens-20M, higher R@100 and NDCG@100 with tied R@10 and MAP for v-BACH on
Taobao, and higher displayed metrics for v-BACH on Netflix. Because this
comparison changes both the spherical family and its learned concentrations,
it supports only an end-to-end comparison between the two complete variants.

\paragraph{Global-codebook scope.}
The selected global p-BACH configurations use 128 codebook heads on
MovieLens-20M and 256 on Taobao and Netflix, while personalized p-BACH uses at
most 32 (Supplementary Table~\ref{tab:selected-k}). They therefore represent
distinct capacity settings.
The global model still predicts mixture weights from the history. It cannot
rotate a direction for user-specific variation within a shared component. In
exchange, its component scores and top lists are user independent.
The current full-catalog protocol characterizes ranking at the reported
capacities.
Supplementary Section~\ref{app:serving} formalizes a residual hybrid between
the two configurations.

\section{Discussion}
\label{sec:discussion}

\paragraph{Choosing a retrieval design.}
For hard multihead retrievers with max-based serving, the controlled results
support per-candidate rather than target-selected routing as the stronger
baseline. Personalized BACH adds a
gate and mixture reranker and reports equal or higher displayed R@10 and MAP
than \texttt{all-multihead} in these comparisons
(Table~\ref{tab:main}). Global BACH trades personalized directions
for reusable lists while remaining competitive with the established baselines
in Table~\ref{tab:main}. They are distinct designs under exact catalog scoring.

\paragraph{Semantic interpretation of latent heads.}
BACH's heads are user-conditioned routing components optimized for ranking, not
predefined semantic categories. The objective has no semantic labels or
disentanglement constraint, is permutation invariant, and allows personalized
directions to vary with history. A personalized head index therefore does not
define a stable topic across users or independent runs. Moreover,
$\pi_u^{(r)}$ is a routing prior, and $\kappa_{ur}$ measures spherical
sharpness; neither variable defines a semantic label.
Retrieved items can be inspected, particularly for a global codebook, but this
alone does not establish semantic coherence. The evidence therefore concerns
ranking utility rather than semantic meaning.

\paragraph{Limitations.}
Tables~\ref{tab:main}, \ref{tab:head-matched}, and~\ref{tab:routing} use one
training run per learned cell. Table~\ref{tab:train-only} repeats Standard and
three $k=32$ multi-interest configurations over five seeds, and the
supplementary training-only rerun of Table~\ref{tab:main} uses seed 0. Those
tables fit PureSVD to the complete retained matrix and fix $\tau=0.07$ and $d=64$; the sensitivity sweep covers
only Taobao, and the ANN study covers catalogs of at most 50,000 items.

\section{Conclusion}

BACH models multi-interest two-tower retrieval as a per-user mixture of
contrastive heads. A spherical recognition model gives differentiable routing,
a separate gate predicts user-specific head mass, and the corresponding
spherical mixture supplies a tractable serving score. The global-codebook
form makes retrieval lists precomputable. A controlled ablation identifies
per-candidate hard routing as a stronger baseline than target-only routing
across every reported head count and dataset. Against this architecture-matched
control, personalized BACH improves R@100 and MAP at every head count and, at
$k=32$, has higher five-seed means under training-only item geometry. It
matches or improves the displayed metrics of the other compared methods under
the shared item geometry and preprocessing.

\section{Ethical Considerations}

Multi-interest recommenders can broaden a candidate set, but they can also
amplify behavioral profiling. The mixture weights in BACH summarize inferred
routing patterns from interaction histories; a deployed system should treat
these weights as sensitive derived data, apply access controls and retention
limits, and avoid presenting latent heads as definitive interests or
explanations about a person. A global codebook is easier to inspect, but it
neither establishes semantic meaning nor replaces privacy review.

Deployed recommenders can create feedback loops that homogenize behavior and
alter item consumption \cite{chaney2018algorithmic}. Deployment should therefore
monitor provider exposure, catalog coverage, subgroup outcomes where lawful and
appropriate, and user controls for correcting or disabling personalization.
Mixture weights should not be used to infer protected traits or make high-impact
decisions.

The experiments use public research datasets and inherit their collection
biases and historical feedback. We report aggregate retrieval metrics for these
benchmark populations.

\bibliographystyle{ACM-Reference-Format}
\bibliography{main}


\begin{thebibliography}{32}


\ifx \showCODEN    \undefined \def \showCODEN     #1{\unskip}     \fi
\ifx \showISBNx    \undefined \def \showISBNx     #1{\unskip}     \fi
\ifx \showISBNxiii \undefined \def \showISBNxiii  #1{\unskip}     \fi
\ifx \showISSN     \undefined \def \showISSN      #1{\unskip}     \fi
\ifx \showLCCN     \undefined \def \showLCCN      #1{\unskip}     \fi
\ifx \shownote     \undefined \def \shownote      #1{#1}          \fi
\ifx \showarticletitle \undefined \def \showarticletitle #1{#1}   \fi
\ifx \showURL      \undefined \def \showURL       {\relax}        \fi
\providecommand\bibfield[2]{#2}
\providecommand\bibinfo[2]{#2}
\providecommand\natexlab[1]{#1}
\providecommand\showeprint[2][]{arXiv:#2}
\makeatletter
\@ifundefined{NAT@parse@date}{}{\let\NAT@parse@date@orig\NAT@parse@date}
\@ifundefined{NAT@parse@date}{}{\def\NAT@parse@date#1#2#3#4#5#6@@{\NAT@parse@date@orig#1#2#3#4#5#6@@\def\NAT@tempyear{0000}\def\NAT@tempexlab{{?}}\ifx\NAT@year\NAT@tempyear\ifx\NAT@exlab\NAT@tempexlab\def\NAT@date{[n.\,d.]}\else\edef\NAT@date{[n.\,d.]\NAT@exlab}\fi\fi}}
\makeatother

\bibitem[Badreddine et~al\mbox{.}(2026)]%
        {badreddine2026theoretical}
\bibfield{author}{\bibinfo{person}{Samy Badreddine}, \bibinfo{person}{Emile
  Van~Krieken}, {and} \bibinfo{person}{Luciano Serafini}.}
  \bibinfo{year}{2026}\natexlab{}.
\newblock \showarticletitle{On the Theoretical Limitations of Embedding-based
  Link Prediction}. In \bibinfo{booktitle}{\emph{Proceedings of the 43rd
  International Conference on Machine Learning}}
  \emph{(\bibinfo{series}{Proceedings of Machine Learning Research},
  Vol.~\bibinfo{volume}{306})}. \bibinfo{publisher}{PMLR},
  \bibinfo{pages}{4866--4893}.
\newblock


\bibitem[Banerjee et~al\mbox{.}(2005)]%
        {banerjee2005clustering}
\bibfield{author}{\bibinfo{person}{Arindam Banerjee},
  \bibinfo{person}{Inderjit~S. Dhillon}, \bibinfo{person}{Joydeep Ghosh}, {and}
  \bibinfo{person}{Suvrit Sra}.} \bibinfo{year}{2005}\natexlab{}.
\newblock \showarticletitle{Clustering on the unit hypersphere using von
  {Mises}--{Fisher} distributions}.
\newblock \bibinfo{journal}{\emph{Journal of Machine Learning Research}}
  \bibinfo{volume}{6} (\bibinfo{year}{2005}), \bibinfo{pages}{1345--1382}.
\newblock


\bibitem[Bennett and Lanning(2007)]%
        {bennett2007netflix}
\bibfield{author}{\bibinfo{person}{James Bennett} {and} \bibinfo{person}{Stan
  Lanning}.} \bibinfo{year}{2007}\natexlab{}.
\newblock \showarticletitle{The {Netflix} Prize}. In
  \bibinfo{booktitle}{\emph{Proceedings of KDD Cup and Workshop}}.
  \bibinfo{pages}{3--6}.
\newblock


\bibitem[Blei et~al\mbox{.}(2017)]%
        {blei2017variational}
\bibfield{author}{\bibinfo{person}{David~M. Blei}, \bibinfo{person}{Alp
  Kucukelbir}, {and} \bibinfo{person}{Jon~D. McAuliffe}.}
  \bibinfo{year}{2017}\natexlab{}.
\newblock \showarticletitle{Variational Inference: A Review for Statisticians}.
\newblock \bibinfo{journal}{\emph{J. Amer. Statist. Assoc.}}
  \bibinfo{volume}{112}, \bibinfo{number}{518} (\bibinfo{year}{2017}),
  \bibinfo{pages}{859--877}.
\newblock


\bibitem[Carbonell and Goldstein(1998)]%
        {carbonell1998use}
\bibfield{author}{\bibinfo{person}{Jaime Carbonell} {and} \bibinfo{person}{Jade
  Goldstein}.} \bibinfo{year}{1998}\natexlab{}.
\newblock \showarticletitle{The use of {MMR}, diversity-based reranking for
  reordering documents and producing summaries}. In
  \bibinfo{booktitle}{\emph{Proceedings of the 21st Annual International ACM
  SIGIR Conference on Research and Development in Information Retrieval}}.
  \bibinfo{pages}{335--336}.
\newblock


\bibitem[Cen et~al\mbox{.}(2020)]%
        {cen2020controllable}
\bibfield{author}{\bibinfo{person}{Yukuo Cen}, \bibinfo{person}{Jianwei Zhang},
  \bibinfo{person}{Xu Zou}, \bibinfo{person}{Chang Zhou},
  \bibinfo{person}{Hongxia Yang}, {and} \bibinfo{person}{Jie Tang}.}
  \bibinfo{year}{2020}\natexlab{}.
\newblock \showarticletitle{Controllable multi-interest framework for
  recommendation}. In \bibinfo{booktitle}{\emph{Proceedings of the 26th ACM
  SIGKDD International Conference on Knowledge Discovery \& Data Mining}}.
  \bibinfo{pages}{2942--2951}.
\newblock


\bibitem[Chaney et~al\mbox{.}(2018)]%
        {chaney2018algorithmic}
\bibfield{author}{\bibinfo{person}{Allison J.~B. Chaney},
  \bibinfo{person}{Brandon~M. Stewart}, {and} \bibinfo{person}{Barbara~E.
  Engelhardt}.} \bibinfo{year}{2018}\natexlab{}.
\newblock \showarticletitle{How Algorithmic Confounding in Recommendation
  Systems Increases Homogeneity and Decreases Utility}. In
  \bibinfo{booktitle}{\emph{Proceedings of the 12th ACM Conference on
  Recommender Systems}}. \bibinfo{pages}{224--232}.
\newblock


\bibitem[Covington et~al\mbox{.}(2016)]%
        {covington2016deep}
\bibfield{author}{\bibinfo{person}{Paul Covington}, \bibinfo{person}{Jay
  Adams}, {and} \bibinfo{person}{Emre Sargin}.}
  \bibinfo{year}{2016}\natexlab{}.
\newblock \showarticletitle{Deep neural networks for {YouTube}
  recommendations}. In \bibinfo{booktitle}{\emph{Proceedings of the 10th ACM
  Conference on Recommender Systems}}. \bibinfo{pages}{191--198}.
\newblock


\bibitem[Cremonesi et~al\mbox{.}(2010)]%
        {cremonesi2010performance}
\bibfield{author}{\bibinfo{person}{Paolo Cremonesi}, \bibinfo{person}{Yehuda
  Koren}, {and} \bibinfo{person}{Roberto Turrin}.}
  \bibinfo{year}{2010}\natexlab{}.
\newblock \showarticletitle{Performance of recommender algorithms on top-n
  recommendation tasks}. In \bibinfo{booktitle}{\emph{Proceedings of the 4th
  ACM Conference on Recommender Systems}}. \bibinfo{pages}{39--46}.
\newblock


\bibitem[Davidson et~al\mbox{.}(2018)]%
        {davidson2018hyperspherical}
\bibfield{author}{\bibinfo{person}{Tim~R. Davidson}, \bibinfo{person}{Luca
  Falorsi}, \bibinfo{person}{Nicola De~Cao}, \bibinfo{person}{Thomas Kipf},
  {and} \bibinfo{person}{Jakub~M. Tomczak}.} \bibinfo{year}{2018}\natexlab{}.
\newblock \showarticletitle{Hyperspherical Variational Auto-Encoders}. In
  \bibinfo{booktitle}{\emph{Proceedings of the 34th Conference on Uncertainty
  in Artificial Intelligence (UAI)}} \emph{(\bibinfo{series}{Proceedings of
  Machine Learning Research}, Vol.~\bibinfo{volume}{R16})}.
  \bibinfo{publisher}{PMLR}, \bibinfo{pages}{855--864}.
\newblock


\bibitem[De~Cao and Aziz(2020)]%
        {decao2020power}
\bibfield{author}{\bibinfo{person}{Nicola De~Cao} {and} \bibinfo{person}{Wilker
  Aziz}.} \bibinfo{year}{2020}\natexlab{}.
\newblock \showarticletitle{The Power Spherical distribution}. In
  \bibinfo{booktitle}{\emph{Proceedings of the 37th International Conference on
  Machine Learning, INNF+ Workshop}}.
\newblock


\bibitem[Harper and Konstan(2015)]%
        {harper2015movielens}
\bibfield{author}{\bibinfo{person}{F.~Maxwell Harper} {and}
  \bibinfo{person}{Joseph~A. Konstan}.} \bibinfo{year}{2015}\natexlab{}.
\newblock \showarticletitle{The {MovieLens} datasets: History and context}.
\newblock \bibinfo{journal}{\emph{ACM Transactions on Interactive Intelligent
  Systems}} \bibinfo{volume}{5}, \bibinfo{number}{4} (\bibinfo{year}{2015}),
  \bibinfo{pages}{1--19}.
\newblock


\bibitem[Hendrycks and Gimpel(2016)]%
        {hendrycks2016gelu}
\bibfield{author}{\bibinfo{person}{Dan Hendrycks} {and} \bibinfo{person}{Kevin
  Gimpel}.} \bibinfo{year}{2016}\natexlab{}.
\newblock \showarticletitle{Gaussian Error Linear Units ({GELU}s)}.
\newblock \bibinfo{journal}{\emph{arXiv preprint arXiv:1606.08415}}
  (\bibinfo{year}{2016}).
\newblock


\bibitem[J{\"a}rvelin and Kek{\"a}l{\"a}inen(2002)]%
        {jarvelin2002cumulated}
\bibfield{author}{\bibinfo{person}{Kalervo J{\"a}rvelin} {and}
  \bibinfo{person}{Jaana Kek{\"a}l{\"a}inen}.} \bibinfo{year}{2002}\natexlab{}.
\newblock \showarticletitle{Cumulated Gain-Based Evaluation of {IR}
  Techniques}.
\newblock \bibinfo{journal}{\emph{ACM Transactions on Information Systems}}
  \bibinfo{volume}{20}, \bibinfo{number}{4} (\bibinfo{year}{2002}),
  \bibinfo{pages}{422--446}.
\newblock


\bibitem[Kang and McAuley(2018)]%
        {kang2018self}
\bibfield{author}{\bibinfo{person}{Wang-Cheng Kang} {and}
  \bibinfo{person}{Julian McAuley}.} \bibinfo{year}{2018}\natexlab{}.
\newblock \showarticletitle{Self-attentive sequential recommendation}. In
  \bibinfo{booktitle}{\emph{2018 IEEE International Conference on Data Mining
  (ICDM)}}. \bibinfo{pages}{197--206}.
\newblock


\bibitem[Karpukhin et~al\mbox{.}(2020)]%
        {karpukhin2020dense}
\bibfield{author}{\bibinfo{person}{Vladimir Karpukhin}, \bibinfo{person}{Barlas
  O{\u{g}}uz}, \bibinfo{person}{Sewon Min}, \bibinfo{person}{Patrick Lewis},
  \bibinfo{person}{Ledell Wu}, \bibinfo{person}{Sergey Edunov},
  \bibinfo{person}{Danqi Chen}, {and} \bibinfo{person}{Wen-tau Yih}.}
  \bibinfo{year}{2020}\natexlab{}.
\newblock \showarticletitle{Dense passage retrieval for open-domain question
  answering}. In \bibinfo{booktitle}{\emph{Proceedings of the 2020 Conference
  on Empirical Methods in Natural Language Processing}}.
  \bibinfo{pages}{6769--6781}.
\newblock


\bibitem[Khattab and Zaharia(2020)]%
        {khattab2020colbert}
\bibfield{author}{\bibinfo{person}{Omar Khattab} {and} \bibinfo{person}{Matei
  Zaharia}.} \bibinfo{year}{2020}\natexlab{}.
\newblock \showarticletitle{{ColBERT}: Efficient and effective passage search
  via contextualized late interaction over {BERT}}. In
  \bibinfo{booktitle}{\emph{Proceedings of the 43rd International ACM SIGIR
  Conference on Research and Development in Information Retrieval}}.
  \bibinfo{pages}{39--48}.
\newblock


\bibitem[Kingma and Ba(2015)]%
        {kingma2015adam}
\bibfield{author}{\bibinfo{person}{Diederik~P. Kingma} {and}
  \bibinfo{person}{Jimmy Ba}.} \bibinfo{year}{2015}\natexlab{}.
\newblock \showarticletitle{Adam: A Method for Stochastic Optimization}. In
  \bibinfo{booktitle}{\emph{International Conference on Learning
  Representations}}.
\newblock


\bibitem[Kulesza and Taskar(2012)]%
        {kulesza2012determinantal}
\bibfield{author}{\bibinfo{person}{Alex Kulesza} {and} \bibinfo{person}{Ben
  Taskar}.} \bibinfo{year}{2012}\natexlab{}.
\newblock \showarticletitle{Determinantal point processes for machine
  learning}.
\newblock \bibinfo{journal}{\emph{Foundations and Trends in Machine Learning}}
  \bibinfo{volume}{5}, \bibinfo{number}{2--3} (\bibinfo{year}{2012}),
  \bibinfo{pages}{123--286}.
\newblock


\bibitem[Li et~al\mbox{.}(2019)]%
        {li2019multi}
\bibfield{author}{\bibinfo{person}{Chao Li}, \bibinfo{person}{Zhiyuan Liu},
  \bibinfo{person}{Mengmeng Wu}, \bibinfo{person}{Yuchi Xu},
  \bibinfo{person}{Huan Zhao}, \bibinfo{person}{Pipei Huang},
  \bibinfo{person}{Guoliang Kang}, \bibinfo{person}{Qiwei Chen},
  \bibinfo{person}{Wei Li}, {and} \bibinfo{person}{Dik~Lun Lee}.}
  \bibinfo{year}{2019}\natexlab{}.
\newblock \showarticletitle{Multi-interest network with dynamic routing for
  recommendation at {Tmall}}. In \bibinfo{booktitle}{\emph{Proceedings of the
  28th ACM International Conference on Information and Knowledge Management}}.
  \bibinfo{pages}{2615--2623}.
\newblock


\bibitem[Liu et~al\mbox{.}(2020)]%
        {liu2020octopus}
\bibfield{author}{\bibinfo{person}{Zheng Liu}, \bibinfo{person}{Jianxun Lian},
  \bibinfo{person}{Junhan Yang}, \bibinfo{person}{Defu Lian}, {and}
  \bibinfo{person}{Xing Xie}.} \bibinfo{year}{2020}\natexlab{}.
\newblock \showarticletitle{Octopus: Comprehensive and elastic user
  representation for the generation of recommendation candidates}. In
  \bibinfo{booktitle}{\emph{Proceedings of the 43rd International ACM SIGIR
  Conference on Research and Development in Information Retrieval}}.
  \bibinfo{pages}{289--298}.
\newblock


\bibitem[{NIST Digital Library of Mathematical Functions}(2026)]%
        {dlmf1041}
\bibfield{author}{\bibinfo{person}{{NIST Digital Library of Mathematical
  Functions}}.} \bibinfo{year}{2026}\natexlab{}.
\newblock \bibinfo{title}{Modified Bessel Functions: Asymptotic Expansions for
  Large Order}.
\newblock \bibinfo{howpublished}{\url{https://dlmf.nist.gov/10.41}}.
\newblock
\newblock
\shownote{Section 10.41, accessed 2026-08-20}.


\bibitem[Pal et~al\mbox{.}(2020)]%
        {pal2020pinnersage}
\bibfield{author}{\bibinfo{person}{Aditya Pal}, \bibinfo{person}{Chantat
  Eksombatchai}, \bibinfo{person}{Yitong Zhou}, \bibinfo{person}{Bo Zhao},
  \bibinfo{person}{Charles Rosenberg}, {and} \bibinfo{person}{Jure Leskovec}.}
  \bibinfo{year}{2020}\natexlab{}.
\newblock \showarticletitle{{PinnerSage}: Multi-modal user embedding framework
  for recommendations at {Pinterest}}. In \bibinfo{booktitle}{\emph{Proceedings
  of the 26th ACM SIGKDD International Conference on Knowledge Discovery \&
  Data Mining}}. \bibinfo{pages}{2311--2320}.
\newblock


\bibitem[Vaswani et~al\mbox{.}(2017)]%
        {vaswani2017attention}
\bibfield{author}{\bibinfo{person}{Ashish Vaswani}, \bibinfo{person}{Noam
  Shazeer}, \bibinfo{person}{Niki Parmar}, \bibinfo{person}{Jakob Uszkoreit},
  \bibinfo{person}{Llion Jones}, \bibinfo{person}{Aidan~N. Gomez},
  \bibinfo{person}{{\L}ukasz Kaiser}, {and} \bibinfo{person}{Illia
  Polosukhin}.} \bibinfo{year}{2017}\natexlab{}.
\newblock \showarticletitle{Attention Is All You Need}. In
  \bibinfo{booktitle}{\emph{Advances in Neural Information Processing
  Systems}}, Vol.~\bibinfo{volume}{30}. \bibinfo{pages}{5998--6008}.
\newblock


\bibitem[Weston et~al\mbox{.}(2013)]%
        {weston2013nonlinear}
\bibfield{author}{\bibinfo{person}{Jason Weston}, \bibinfo{person}{Ron~J.
  Weiss}, {and} \bibinfo{person}{Hector Yee}.} \bibinfo{year}{2013}\natexlab{}.
\newblock \showarticletitle{Nonlinear latent factorization by embedding
  multiple user interests}. In \bibinfo{booktitle}{\emph{Proceedings of the 7th
  ACM Conference on Recommender Systems}}. \bibinfo{pages}{65--68}.
\newblock


\bibitem[Xie et~al\mbox{.}(2023)]%
        {xie2023rethinking}
\bibfield{author}{\bibinfo{person}{Yueqi Xie}, \bibinfo{person}{Jingqi Gao},
  \bibinfo{person}{Peilin Zhou}, \bibinfo{person}{Qichen Ye},
  \bibinfo{person}{Yining Hua}, \bibinfo{person}{Jae~Boum Kim},
  \bibinfo{person}{Fangzhao Wu}, {and} \bibinfo{person}{Sunghun Kim}.}
  \bibinfo{year}{2023}\natexlab{}.
\newblock \showarticletitle{Rethinking multi-interest learning for candidate
  matching in recommender systems}. In \bibinfo{booktitle}{\emph{Proceedings of
  the 17th ACM Conference on Recommender Systems}}. \bibinfo{pages}{283--293}.
\newblock


\bibitem[Xiong et~al\mbox{.}(2020)]%
        {xiong2020layer}
\bibfield{author}{\bibinfo{person}{Ruibin Xiong}, \bibinfo{person}{Yunchang
  Yang}, \bibinfo{person}{Di He}, \bibinfo{person}{Kai Zheng},
  \bibinfo{person}{Shuxin Zheng}, \bibinfo{person}{Chen Xing},
  \bibinfo{person}{Huishuai Zhang}, \bibinfo{person}{Yanyan Lan},
  \bibinfo{person}{Liwei Wang}, {and} \bibinfo{person}{Tie-Yan Liu}.}
  \bibinfo{year}{2020}\natexlab{}.
\newblock \showarticletitle{On Layer Normalization in the Transformer
  Architecture}. In \bibinfo{booktitle}{\emph{Proceedings of the 37th
  International Conference on Machine Learning}}
  \emph{(\bibinfo{series}{Proceedings of Machine Learning Research},
  Vol.~\bibinfo{volume}{119})}. \bibinfo{publisher}{PMLR},
  \bibinfo{pages}{10524--10533}.
\newblock


\bibitem[Yang et~al\mbox{.}(2020)]%
        {yang2020mixed}
\bibfield{author}{\bibinfo{person}{Ji Yang}, \bibinfo{person}{Xinyang Yi},
  \bibinfo{person}{Derek~Zhiyuan Cheng}, \bibinfo{person}{Lichan Hong},
  \bibinfo{person}{Yang Li}, \bibinfo{person}{Simon~Xiaoming Wang},
  \bibinfo{person}{Taibai Xu}, {and} \bibinfo{person}{Ed~H. Chi}.}
  \bibinfo{year}{2020}\natexlab{}.
\newblock \showarticletitle{Mixed negative sampling for learning two-tower
  neural networks in recommendations}. In \bibinfo{booktitle}{\emph{Companion
  Proceedings of the Web Conference 2020}}. \bibinfo{pages}{441--447}.
\newblock


\bibitem[Yang et~al\mbox{.}(2018)]%
        {yang2018breaking}
\bibfield{author}{\bibinfo{person}{Zhilin Yang}, \bibinfo{person}{Zihang Dai},
  \bibinfo{person}{Ruslan Salakhutdinov}, {and} \bibinfo{person}{William~W.
  Cohen}.} \bibinfo{year}{2018}\natexlab{}.
\newblock \showarticletitle{Breaking the softmax bottleneck: A high-rank {RNN}
  language model}. In \bibinfo{booktitle}{\emph{International Conference on
  Learning Representations}}.
\newblock


\bibitem[Yi et~al\mbox{.}(2019)]%
        {yi2019sampling}
\bibfield{author}{\bibinfo{person}{Xinyang Yi}, \bibinfo{person}{Ji Yang},
  \bibinfo{person}{Lichan Hong}, \bibinfo{person}{Derek~Zhiyuan Cheng},
  \bibinfo{person}{Lukasz Heldt}, \bibinfo{person}{Aditee Kumthekar},
  \bibinfo{person}{Zhe Zhao}, \bibinfo{person}{Li Wei}, {and}
  \bibinfo{person}{Ed Chi}.} \bibinfo{year}{2019}\natexlab{}.
\newblock \showarticletitle{Sampling-bias-corrected neural modeling for large
  corpus item recommendations}. In \bibinfo{booktitle}{\emph{Proceedings of the
  13th ACM Conference on Recommender Systems}}. \bibinfo{pages}{269--277}.
\newblock


\bibitem[Zhang et~al\mbox{.}(2023)]%
        {zhang2023mirn}
\bibfield{author}{\bibinfo{person}{Xiliang Zhang}, \bibinfo{person}{Jin Liu},
  \bibinfo{person}{Siwei Chang}, \bibinfo{person}{Peizhu Gong},
  \bibinfo{person}{Zhongdai Wu}, {and} \bibinfo{person}{Bing Han}.}
  \bibinfo{year}{2023}\natexlab{}.
\newblock \showarticletitle{{MIRN}: A multi-interest retrieval network with
  sequence-to-interest {EM} routing}.
\newblock \bibinfo{journal}{\emph{PLOS ONE}} \bibinfo{volume}{18},
  \bibinfo{number}{2} (\bibinfo{year}{2023}), \bibinfo{pages}{e0281275}.
\newblock


\bibitem[Zhu et~al\mbox{.}(2018)]%
        {zhu2018learning}
\bibfield{author}{\bibinfo{person}{Han Zhu}, \bibinfo{person}{Xiang Li},
  \bibinfo{person}{Pengye Zhang}, \bibinfo{person}{Guozheng Li},
  \bibinfo{person}{Jie He}, \bibinfo{person}{Han Li}, {and}
  \bibinfo{person}{Kun Gai}.} \bibinfo{year}{2018}\natexlab{}.
\newblock \showarticletitle{Learning tree-based deep model for recommender
  systems}. In \bibinfo{booktitle}{\emph{Proceedings of the 24th ACM SIGKDD
  International Conference on Knowledge Discovery \& Data Mining}}.
  \bibinfo{pages}{1079--1088}.
\newblock


\end{thebibliography}

\clearpage
\appendix
\section*{Supplementary Material}
\section{Experimental Details}
\label{app:implementation}

\paragraph{Splits and full-catalog evaluation.}
For a sequence $(i_1,\ldots,i_T)$, items $i_{T-19:T-10}$ are validation
targets and $i_{T-9:T}$ are test targets. The validation history is the last 50
items before the validation segment; the test history additionally includes
that segment. Training windows stop before both held-out segments. Each
stride-one training window contains 50 history items and samples one target
from the next ten positions. We evaluate future-item retrieval by scoring every
eligible catalog item for a sample of at most 10,000 qualifying users.
The protocol uses interaction-level holdouts and exact full-catalog evaluation.
Table~\ref{tab:data} reports the resulting dataset sizes and training-window
counts.

\begin{table}[H]
\centering
\caption{Processed datasets. ``Windows'' is the number of training windows.}
\label{tab:data}
\small
\setlength{\tabcolsep}{4pt}
\begin{tabular}{lrrr}
\toprule
Dataset & Users & Items & Windows \\
\midrule
MovieLens-20M & 36,648 & 20,720 & 3,957,331 \\
Taobao        & 69,617 & 50,000 & 2,164,022 \\
Netflix       & 50,000 & 17,769 & 8,249,567 \\
\bottomrule
\end{tabular}
\end{table}

\paragraph{Shared towers and encoders.}
All learned methods use normalized 64-dimensional vectors and a shared item
tower consisting of a frozen rank-64 PureSVD table
\cite{cremonesi2010performance} followed by a zero-initialized residual
MLP of width 128. Standard, hard multihead, and BACH use
a one-layer Transformer \cite{vaswani2017attention} with
\emph{pre-layer normalization}~(pre-LN) \cite{xiong2020layer}, learned summary
tokens, four
attention heads, feed-forward width 128, and dropout 0.1. Standard prepends one
summary token, applies no causal mask, and uses that token's output as its
single user vector. SASRec instead applies two causal self-attention blocks and
uses the final history position as its single user vector. MIND and ComiRec
retain their capsule or self-attentive extractors.

\paragraph{Model configurations.}
The BACH head tower prepends $k$ learned summary tokens to the history. Its
separate gating tower has the same depth and width, followed by one scalar
projection per summary output and a softmax over heads. The concentration
network is a two-hidden-layer \emph{Gaussian error linear unit}~(GELU) MLP
\cite{hendrycks2016gelu} of width $4d=256$, shared across heads. Its final
layer is zero initialized with bias $\log(1/\tau)$, so all concentrations start
at $1/\tau$.

MIND and ComiRec-DR use three capsule-routing iterations;
ComiRec-SA uses self-attentive extraction; MIND uses label-aware attention; and
both hard controls use the BACH summary-token encoder. Thus objectives differ
by method while the item tower, candidate sampler, optimizer, and evaluation
are matched. MostPop ranks items by training-set frequency for every user after
excluding that user's request-history items.

\paragraph{Sampling and optimization.}
Random negatives are resampled once per minibatch and shared across examples.
The other 511 batch positives and 2,048 random draws form the remaining
candidate columns; duplicate item IDs are retained. Adam uses its default
momentum coefficients with no weight decay, learning-rate schedule, or gradient
clipping. Validation MAP is checked every five epochs with patience ten, and
the best checkpoint is restored.
Tables~\ref{tab:configuration} and~\ref{tab:selected-k} summarize the shared
training setup and the validation-selected head counts used for
Table~\ref{tab:main}.

\begin{table}[H]
\centering
\caption{Shared training configuration.}
\label{tab:configuration}
\small
\setlength{\tabcolsep}{4pt}
\begin{tabular}{ll}
\toprule
Setting & Value \\
\midrule
Embedding / history & $d=64$ / 50 items \\
Summary-token Transformer & 1 layer, 4 heads, feed-forward 128 \\
Item adaptation & residual MLP 128, zero-init \\
Normalization / temperature & $\ell_2$ / fixed $\tau=.07$ \\
Optimizer & Adam, learning rate $10^{-3}$ \\
Batch / random negatives & 512 / 2,048 \\
Early stopping & validation MAP, patience 10 \\
Evaluation / epoch cap & every 5 epochs / 200 (Taobao: 50) \\
Retrieval heads & $k\in\{8,16,32\}$ \\
Global codebook heads & $k\in\{128,256\}$ \\
Hardware & 1 NVIDIA A10G \\
\bottomrule
\end{tabular}
\end{table}

\begin{table}[H]
\centering
\caption{Head counts selected by validation MAP for Table~\ref{tab:main}.}
\label{tab:selected-k}
\small
\setlength{\tabcolsep}{3.6pt}
\begin{tabular}{lccc}
\toprule
Family & ML20M & Taobao & Netflix \\
\midrule
MIND          & 16 & 8  & 16 \\
ComiRec-DR    & 8  & 8  & 8 \\
ComiRec-SA    & 8  & 8  & 8 \\
\texttt{pos-multihead} & 8  & 8  & 8 \\
\texttt{all-multihead} & 16 & 16 & 32 \\
p-BACH        & 32 & 32 & 32 \\
v-BACH        & 32 & 32 & 32 \\
global p-BACH & 128 & 256 & 256 \\
\bottomrule
\end{tabular}
\end{table}

\section{Derivations and Numerical Evaluation}
\label{app:derivations}

\subsection{Spherical Normalizers}

For the power-spherical density, let
$\alpha=(d-1)/2+\kappa$ and $\beta=(d-1)/2$. Its normalizer is
\begin{equation}
C_d^{\mathrm{PS}}(\kappa)
=\frac{\Gamma(\alpha+\beta)}
{2^{\alpha+\beta}\pi^\beta\Gamma(\alpha)}.
\label{eq:ps-normalizer}
\end{equation}
For vMF the normalizer is
\begin{equation}
C_d^{\mathrm{vMF}}(\kappa)
=\frac{\kappa^{d/2-1}}
{(2\pi)^{d/2}I_{d/2-1}(\kappa)},
\label{eq:vmf-normalizer}
\end{equation}
where $I_\nu$ is a modified Bessel function. The power-spherical kernel is
Bessel-free, whereas vMF uses an exponential directional density. In
both cases the normalizer rescales each component. Omitting it when
concentrations vary across heads would compare scores measured on different
scales.

\subsection{Optimization Caveats}

The marginal $\ell$ in Equation~\eqref{eq:marginal} does not contain
concentration. Concentration parameterizes only the recognition distribution
and hence the ELBO gap in Equation~\eqref{eq:gap}; for fixed generative
parameters, optimizing it minimizes that gap. This permits bounded values
without an explicit prior but does not guarantee a non-degenerate optimum. The
gate appears both explicitly in the prior term and implicitly in
Equation~\eqref{eq:gamma}; we preserve both paths and train it by ordinary
backpropagation. Thus $\pi_u$ is an amortized latent-component prior rather than
a closed-form frequency estimate. Although the responsibility formula does not
depend on the sampled candidate multiset for fixed parameters, the learned gate
is influenced by Equation~\eqref{eq:component}. It thus represents sampled
contrastive mixture weights.

\subsection{Exchangeability and Stable Evaluation}

The objective is invariant to a common permutation of the heads, gate
coordinates, and concentration outputs. Head index $r$ consequently has no
intrinsic meaning across independent runs. The personalized model can also
change a head's direction with the user, so a coordinate should not be labeled
as a universal topic. The global codebook makes directions shared and easier
to inspect, but its indices remain identifiable only up to permutation.

We compute responsibilities and mixture scores in log space, using log-softmax
for Equation~\eqref{eq:gamma} and log-sum-exp for
Equation~\eqref{eq:smooth-max}. For the power-spherical kernel,
$1+\langle\mbf{u},\mbf{v}\rangle$ is clamped at $10^{-6}$ before taking a
log. For vMF, let $\nu=d/2-1$, $z=\kappa/\nu$,
$s=(1+z^2)^{1/2}$, $\eta=s+\log(z/(1+s))$, $t=1/s$, and
$u_1(t)=(3t-5t^3)/24$. We use the differentiable first-correction term of the
standard large-order expansion \cite{dlmf1041},
\begin{equation}
\log I_\nu(\kappa)\approx
-\tfrac12\log(2\pi\nu)+\nu\eta-\tfrac14\log(1+z^2)
+\log\!\left(1+\frac{u_1(t)}{\nu}\right).
\label{eq:bessel-approx}
\end{equation}

For $d=64$ ($\nu=31$), a double-precision check at 10,000 logarithmically
spaced concentrations in $[10^{-4},30]$ gives a maximum absolute error of
$3.4\times10^{-5}$ in $\log I_\nu(\kappa)$.
The clamp and approximation alter the numerical evaluation of the stated
objective but avoid undefined logarithms and unstable direct Bessel
evaluation. We retain $C_d(\kappa_{ur})$ because concentrations differ across
components, so it is not a common factor that can be removed from the mixture;
it remains item-independent within each component.

\section{Head-Count Sensitivity}
\label{app:head-count}

\begin{figure*}[t]
\centering
\includegraphics[width=0.86\textwidth]{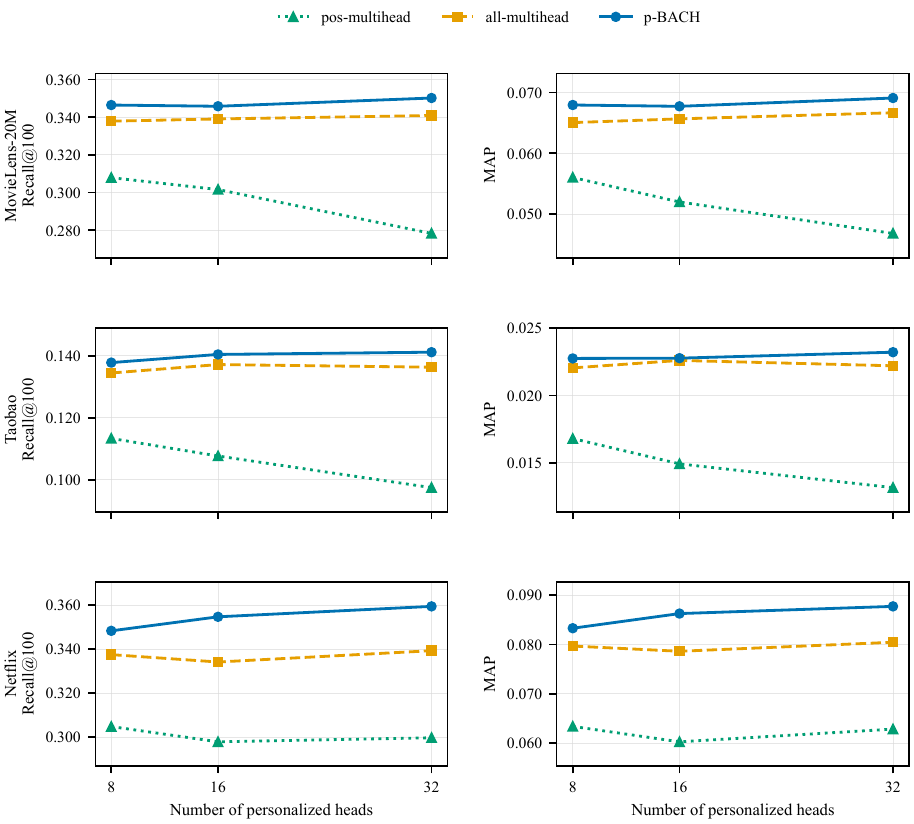}
\caption{Full-catalog R@100 and MAP across personalized head counts
$k\in\{8,16,32\}$. Points are unrounded results, and lines connect
tested configurations without smoothing. Comparisons are head-matched rather
than parameter-matched.}
\Description{Six line plots show Recall at 100 and mean average precision for
MovieLens-20M, Taobao, and Netflix as the number of personalized heads changes
from 8 to 32. Per-candidate \texttt{all-multihead} remains above
target-selected \texttt{pos-multihead}, while p-BACH remains above
\texttt{all-multihead} at every tested head count.}
\label{fig:performance-vs-k}
\end{figure*}

Figure~\ref{fig:performance-vs-k} places the matched routing and p-BACH
comparisons on the same axes. At every tested personalized head count,
per-candidate \texttt{all-multihead} exceeds target-selected
\texttt{pos-multihead}, and p-BACH exceeds \texttt{all-multihead} in unrounded
R@100 and MAP on all three datasets. The lines connect the three observed
points; they do not interpolate untested head counts.

\section{Serving Analysis and Direction-Sharing Variants}
\label{app:serving}

\subsection{Direction-Personalization Spectrum}

\begin{figure*}[t]
\centering
\includegraphics{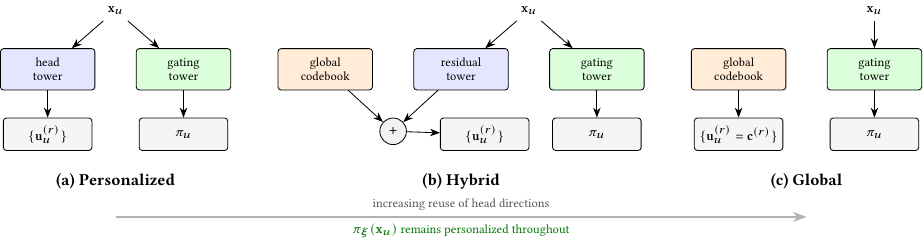}
\caption{BACH's direction-personalization spectrum at the tower level.
Personalized BACH predicts all head directions from the history. A hybrid adds
a user-conditioned residual to a shared codebook and $\ell_2$-normalizes the sum.
Global BACH uses only shared, precomputable directions. The gating tower remains
history-conditioned in all three configurations.}
\Description{Three tower diagrams compare personalized, hybrid, and global
BACH. All receive user history in a gating tower. The personalized model also
feeds history to a head tower, the hybrid combines a global codebook with a
history-conditioned residual, and the global model uses only the shared
codebook for head directions.}
\label{fig:spectrum}
\end{figure*}

Intermediate configurations can combine global codebook heads with
user-conditioned residuals or mix global and personalized heads. Let
$R_\vartheta(\mbf{x}_u)=
(\bm{\delta}_u^{(1)},\ldots,\bm{\delta}_u^{(k)})\in\mbb{R}^{k\times d}$ be the
residual-tower output, and define
$\operatorname{normalize}(\mbf{a})=\mbf{a}/\lVert\mbf{a}\rVert_2$. The residual
variant uses
$\mbf{u}_u^{(r)}=\operatorname{normalize}
(\mbf{c}^{(r)}+\bm{\delta}_u^{(r)})$
in the component likelihood, recognition responsibilities, and serving score.
These variants interpolate retrieval cost and expressiveness without changing
the objective. Figure~\ref{fig:spectrum} shows this residual hybrid.

\section{Additional Analyses}
\label{app:additional}

\subsection{Training-Only Item Geometry}
\label{app:train-only}

The robustness check in Section~\ref{sec:results} refits PureSVD after removing
the final 20 interactions (ten validation and ten test items) of every user
eligible for evaluation, while retaining the original cohort and catalog. The
same 10,000 users are evaluated for every method and seed. Seeds 0--4 change the
parameter initialization, the order of training windows, and the sampled
negatives. The refit leaves 1,319, 143, and 3 catalog items with zero PureSVD
rows on MovieLens-20M, Netflix, and Taobao, and these items account for 0.18\%,
0.02\%, and 0\% of test targets, respectively.

Table~\ref{tab:train-only-main} reruns the selected configurations of
Table~\ref{tab:main} with this geometry and seed 0. Netflix global p-BACH
($k=256$) was not rerun because one run takes about 59 GPU-hours. Personalized
BACH variants attain the highest displayed R@100 and MAP on every dataset among
the rerun methods. SASRec has the same displayed R@100 as \texttt{all-multihead}
on MovieLens-20M and a higher value on Netflix.

\begin{table*}[t]
\centering
\caption{Selected configurations of Table~\ref{tab:main} retrained with
training-only PureSVD and seed 0. Columns as in Table~\ref{tab:main}; bold
marks the best displayed value in each dataset and metric.}
\label{tab:train-only-main}
\scriptsize
\setlength{\tabcolsep}{3.1pt}
\begin{tabular}{lcccccccccccc}
\toprule
& \multicolumn{4}{c}{ML20M} & \multicolumn{4}{c}{Taobao} & \multicolumn{4}{c}{Netflix} \\
\cmidrule(lr){2-5}\cmidrule(lr){6-9}\cmidrule(lr){10-13}
Method & R10 & R100 & N100 & MAP & R10 & R100 & N100 & MAP & R10 & R100 & N100 & MAP \\
\midrule
MIND & .024 & .165 & .082 & .025 & .009 & .059 & .030 & .008 & .019 & .124 & .064 & .019 \\
ComiRec-DR & .031 & .194 & .100 & .031 & .010 & .066 & .034 & .009 & .050 & .230 & .130 & .042 \\
ComiRec-SA & .047 & .261 & .140 & .045 & .017 & .095 & .051 & .013 & .060 & .274 & .156 & .052 \\
\texttt{pos-multihead} & .056 & .293 & .158 & .052 & .018 & .104 & .055 & .014 & .087 & .322 & .196 & .073 \\
Standard & .065 & .328 & .180 & .062 & .020 & .107 & .058 & .016 & .089 & .327 & .201 & .076 \\
SASRec & .067 & .333 & .184 & .064 & .020 & .107 & .058 & .016 & .096 & .350 & .215 & .081 \\
\texttt{all-multihead} & .068 & .333 & .184 & .064 & .024 & .120 & .066 & .018 & .095 & .334 & .208 & .080 \\
p-BACH & .072 & .344 & .192 & \textbf{.068} & \textbf{.025} & \textbf{.123} & \textbf{.068} & \textbf{.019} & \textbf{.105} & \textbf{.360} & \textbf{.227} & \textbf{.089} \\
v-BACH & \textbf{.073} & \textbf{.347} & \textbf{.194} & \textbf{.068} & \textbf{.025} & \textbf{.123} & \textbf{.068} & \textbf{.019} & .100 & .353 & .221 & .086 \\
global p-BACH & .054 & .299 & .159 & .053 & .016 & .098 & .051 & .014 & -- & -- & -- & -- \\
\bottomrule
\end{tabular}
\end{table*}

\subsection{Recall and NDCG Pareto Set}
\label{app:pareto}

\begin{figure*}[t]
\centering
\includegraphics{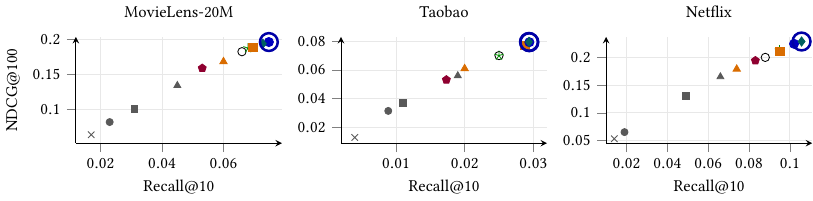}
\caption{Recall@10 versus NDCG@100 for every method in
Table~\ref{tab:main}. Higher values are better on both axes. Blue outlines and
the connecting segment mark the non-dominated set using the unrounded values
underlying the table.}
\Description{Three scatter plots compare Recall at 10 with NDCG at 100 on
MovieLens-20M, Taobao, and Netflix. Personalized BACH variants occupy each
dataset's upper-right Pareto set. The Taobao frontier contains both p-BACH and
v-BACH; the other two frontiers contain one BACH variant each.}
\label{fig:pareto-metrics}
\end{figure*}

Figure~\ref{fig:pareto-metrics} plots Recall@10 against NDCG@100. This pair
contrasts short-list coverage with rank-sensitive quality through depth 100
for every method in Table~\ref{tab:main}. A method is non-dominated
if no other method is at least as high on both axes and strictly higher on one.
At unrounded precision, the Pareto set contains p-BACH on MovieLens-20M,
p-BACH and v-BACH on Taobao, and v-BACH on Netflix. The Taobao pair exhibits
the only trade-off: p-BACH has higher Recall@10 (.02942 versus .02936), whereas
v-BACH has higher NDCG@100 (.07952 versus .07932). Because the axes remain
nearly monotone across methods and the other two frontiers each reduce to one
point, Figure~\ref{fig:pareto-metrics} primarily summarizes dominance, with a
measured trade-off only on Taobao.

\subsection{Serving Rules}
\label{app:serving-rules}

Table~\ref{tab:serving-rules} rescores the $k=32$ checkpoints of
Table~\ref{tab:head-matched} with exchanged serving rules and no retraining.
Let $s_{ur}(i)=\log\pi_u^{(r)}+\log\sigma_{ur}(\mbf{v}_i)$; the largest
mixture term is $\max_r s_{ur}(i)$. The uniform soft maximum
\[
\log\textstyle\sum_r\exp\bigl(\langle\mbf{u}_u^{(r)},\mbf{v}_i\rangle/\tau\bigr)
\]
induces the same item ranking as a uniform vMF mixture with concentration
$1/\tau$. The catalog-normalized
mixture is $\log\sum_r\pi_u^{(r)}p_r(i\mid u)$, where each component softmax of
Equation~\eqref{eq:component} is normalized over the full catalog; it needs
$k$ catalog normalizers per request, and \texttt{all-multihead} uses
$\pi_u^{(r)}=1/k$. Scores are computed on a CPU, so the default rows match
Table~\ref{tab:head-matched} up to $3\times10^{-4}$.

\begin{table}[H]
\centering
\caption{Exchanged serving rules for the $k=32$ checkpoints (R@100/MAP). The
first row of each block uses the model's default serving rule.}
\label{tab:serving-rules}
\scriptsize
\setlength{\tabcolsep}{2pt}
\begin{tabular}{llccc}
\toprule
Trained & Served by & MovieLens-20M & Taobao & Netflix \\
\midrule
p-BACH & spherical mixture (Eq.~\ref{eq:serving}) & .350/.069 & .141/.023 & .359/.088 \\
 & largest mixture term & .339/.068 & .129/.021 & .340/.081 \\
 & mixture with $\pi_u^{(r)}=1/k$ & .320/.059 & .133/.021 & .303/.063 \\
 & unweighted maximum & .286/.051 & .111/.017 & .291/.057 \\
 & catalog-normalized softmax mixture & .357/.071 & .141/.023 & .367/.090 \\
\midrule
\texttt{all-multihead} & unweighted maximum & .341/.067 & .136/.022 & .339/.080 \\
 & uniform soft maximum & .335/.064 & .134/.021 & .334/.077 \\
 & catalog-normalized softmax mixture & .328/.061 & .134/.021 & .328/.076 \\
\bottomrule
\end{tabular}
\end{table}

\subsection{Diversity and Long-Tail Exposure}
\label{app:diversity}

Table~\ref{tab:diversity} summarizes the top-100 lists of the methods in
Table~\ref{tab:main} for the same evaluation users. Distinct items count the
items recommended to at least one user. The tail holds every item outside the
20\% most frequent in training; tail share is the mean fraction of list slots
holding tail items, and tail coverage is the fraction of tail items recommended
at least once. Genre coverage is the mean fraction of a user's held-out
MovieLens-20M genres present in the list.

\begin{table*}[t]
\centering
\caption{Top-100 list diversity (distinct items, tail share, tail coverage) and
MovieLens-20M held-out genre coverage. Bold marks the largest unrounded learned
value.}
\label{tab:diversity}
\scriptsize
\setlength{\tabcolsep}{3pt}
\begin{tabular}{lcccccccccc}
\toprule
& \multicolumn{3}{c}{MovieLens-20M} & \multicolumn{3}{c}{Taobao} & \multicolumn{3}{c}{Netflix} & ML20M \\
\cmidrule(lr){2-4}\cmidrule(lr){5-7}\cmidrule(lr){8-10}\cmidrule(lr){11-11}
Method & Distinct & Tail share & Tail cov. & Distinct & Tail share & Tail cov. & Distinct & Tail share & Tail cov. & Genre cov. \\
\midrule
MostPop & 0.1k & .000 & .000 & 0.1k & .000 & .000 & 0.1k & .000 & .000 & .982 \\
\midrule
MIND & 4.9k & .009 & .065 & 27.2k & .466 & .441 & 5.2k & .023 & .118 & .984 \\
ComiRec-DR & 7.0k & .044 & .173 & 32.7k & .405 & .569 & 7.0k & .082 & .244 & .974 \\
ComiRec-SA & 8.2k & .080 & .245 & 33.2k & .326 & .582 & 7.0k & .063 & .250 & .980 \\
\texttt{pos-multihead} & 7.8k & .045 & .223 & 33.8k & .318 & .599 & 7.4k & .089 & .270 & .980 \\
Standard & 9.0k & .067 & .292 & 39.6k & .421 & .740 & 8.0k & .104 & .311 & .966 \\
SASRec & 9.7k & .076 & .336 & 40.1k & .413 & .754 & 7.9k & .126 & .304 & .965 \\
\texttt{all-multihead} & 9.8k & .080 & .341 & 39.9k & .451 & .748 & 8.3k & .144 & .337 & .968 \\
p-BACH & \textbf{10.1k} & .081 & \textbf{.361} & \textbf{40.3k} & .447 & \textbf{.758} & \textbf{8.8k} & .134 & \textbf{.367} & .970 \\
v-BACH & 10.1k & .087 & .361 & 40.1k & .447 & .752 & 8.7k & .129 & .360 & .967 \\
global p-BACH & 5.1k & .031 & .074 & 21.1k & .281 & .297 & 6.1k & .106 & .188 & .972 \\
\bottomrule
\end{tabular}
\end{table*}

\subsection{Candidate Generation and Latency}
\label{app:ann}

Table~\ref{tab:ann} evaluates the serving pipeline of
Section~\ref{sec:serving} on the $k=32$ checkpoints and the selected global
models. Each head retrieves its top $M$ items from a FAISS HNSW index (graph
degree 32, search breadth $\max(M,128)$). History items are removed from the
union, the rest are reranked by the serving score, and held-out items outside
the union count as misses. Latency is the median per user on one thread of an
Intel Xeon Platinum 8488C CPU, from encoding the user to the final top 100. For global
p-BACH, the lists and component scores are computed once offline, so the online
work is the gate and the reweighting of a larger union. Relative to exact
per-head top-$M$ retrieval, HNSW changes R@100 by at most $4\times10^{-5}$.
Across datasets, the held-out-item recall of the candidate union is
.411--.761, .671--.939, and .773--.971 for p-BACH at $M=100$, 500, and 1000, and
.339--.579, .586--.843, and .691--.922 for \texttt{all-multihead}.

\begin{table*}[t]
\centering
\caption{Per-head HNSW retrieval of $M$ candidates, union, and reranking:
R@100/MAP, mean union size, and median single-thread CPU latency per user (ms).
``full'' scores the entire catalog exactly.}
\label{tab:ann}
\scriptsize
\setlength{\tabcolsep}{3pt}
\begin{tabular}{llccccccccc}
\toprule
& & \multicolumn{3}{c}{MovieLens-20M} & \multicolumn{3}{c}{Taobao} & \multicolumn{3}{c}{Netflix} \\
\cmidrule(lr){3-5}\cmidrule(lr){6-8}\cmidrule(lr){9-11}
Model & $M$ & R@100/MAP & Union & ms & R@100/MAP & Union & ms & R@100/MAP & Union & ms \\
\midrule
\texttt{all-multihead} & full & .341/.067 & -- & -- & .136/.022 & -- & -- & .339/.080 & -- & -- \\
 & 100 & .341/.060 & 0.2k & 3.5 & .136/.020 & 0.6k & 4.7 & .339/.077 & 0.4k & 4.6 \\
 & 500 & .341/.066 & 0.9k & 15.7 & .136/.022 & 2.5k & 21.3 & .339/.080 & 1.4k & 19.8 \\
 & 1000 & .341/.066 & 1.6k & 42.1 & .136/.022 & 4.2k & 52.9 & .339/.080 & 2.3k & 50.9 \\
\midrule
p-BACH & full & .350/.069 & -- & -- & .141/.023 & -- & -- & .359/.088 & -- & -- \\
 & 100 & .350/.067 & 0.6k & 5.8 & .141/.022 & 1.1k & 8.4 & .359/.086 & 0.9k & 6.5 \\
 & 500 & .350/.069 & 1.7k & 20.6 & .141/.023 & 4.0k & 32.3 & .359/.088 & 2.7k & 22.5 \\
 & 1000 & .350/.069 & 2.7k & 51.1 & .141/.023 & 6.8k & 76.0 & .359/.088 & 4.0k & 54.4 \\
\midrule
global p-BACH & full & .297/.053 & -- & -- & .102/.014 & -- & -- & .317/.072 & -- & -- \\
 & 100 & .297/.052 & 4.9k & 5.5 & .102/.014 & 20.6k & 37.7 & .317/.072 & 7.1k & 8.6 \\
 & 500 & .297/.053 & 9.6k & 9.6 & .102/.014 & 41.2k & 94.1 & .317/.072 & 11.8k & 13.8 \\
 & 1000 & .297/.053 & 11.9k & 11.5 & .102/.014 & 46.2k & 121.3 & .317/.072 & 13.8k & 16.4 \\
\bottomrule
\end{tabular}
\end{table*}

\subsection{Temperature and Dimension}
\label{app:sensitivity}

Table~\ref{tab:sensitivity} varies the contrastive temperature $\tau$ and the
embedding dimension $d$ for \texttt{all-multihead} and p-BACH with $k=32$ on
Taobao, one run per setting, under the protocol of Table~\ref{tab:main}. For
p-BACH, $\tau$ also sets the initial concentration $1/\tau$. The dimension
rows use $d$-dimensional PureSVD tables built in the same way, with $\tau=0.07$.

\begin{table}[H]
\centering
\caption{Taobao sensitivity of the $k=32$ models (R@100/MAP). Bold marks the
higher unrounded R@100 in each row.}
\label{tab:sensitivity}
\scriptsize
\begin{tabular}{lcc}
\toprule
Setting & \texttt{all-multihead} & p-BACH \\
\midrule
$\tau=0.03$ & .137/.022 & \textbf{.142/.023} \\
$\tau=0.05$ & .136/.022 & \textbf{.143/.023} \\
$\tau=0.07$, $d=64$ (paper) & .136/.022 & \textbf{.141/.023} \\
$\tau=0.1$ & .136/.022 & \textbf{.138/.022} \\
$\tau=0.2$ & \textbf{.119/.017} & .113/.017 \\
\midrule
$d=32$ & .122/.019 & \textbf{.125/.019} \\
$d=128$ & .151/.026 & \textbf{.156/.026} \\
\bottomrule
\end{tabular}
\end{table}

\subsection{Routing Diagnostics}
\label{app:mechanism}

\begin{figure*}[t]
\centering
\includegraphics[width=0.83\textwidth]{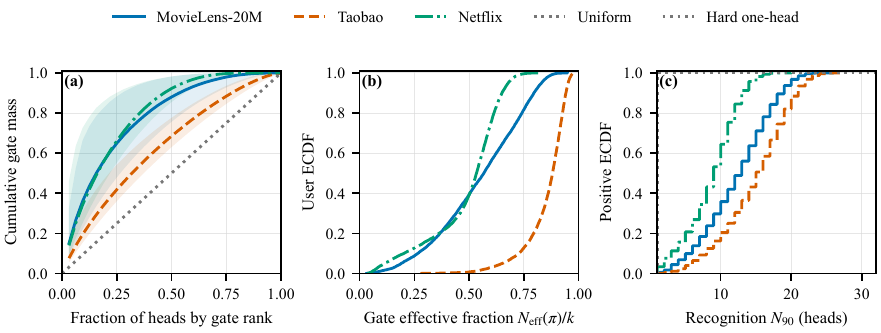}
\caption{Personalized p-BACH routing diagnostics at $k=32$, using 10,000 users
and 100,000 held-out positives per dataset. (a) Heads are sorted within each
user by gate weight; curves show median cumulative mass and shaded 10th--90th
percentiles. (b) User \emph{empirical cumulative distribution function}~(ECDF)
of the
effective head fraction $N_{\mathrm{eff}}(\pi_u)/k$. (c) Held-out-positive ECDF
of $N_{90}$, the minimum number of largest recognition responsibilities whose
sum reaches 90\%. The references show uniform gate mass and one-head hard
routing.}
\Description{Three plots compare personalized p-BACH routing on MovieLens-20M,
Taobao, and Netflix. Ranked cumulative gate-mass curves exceed a uniform
diagonal. Effective gate fractions vary across users, and most held-out
positives require multiple heads to contain 90 percent of recognition
responsibility.}
\label{fig:mechanism}
\end{figure*}

Figure~\ref{fig:mechanism} details the gate and recognition responsibilities
of personalized p-BACH at $k=32$ summarized in
Section~\ref{sec:routing-analysis}. For a distribution $a$, let
$H(a)=-\sum_r a^{(r)}\log a^{(r)}$ be its Shannon entropy and
$N_{\mathrm{eff}}(a)=\exp(H(a))$ its effective count. Heads are sorted within
each user because their indices have no intrinsic cross-user meaning. The
cumulative gate-mass curves lie above the uniform reference on all three
datasets, and the gate is more diffuse on Taobao than on MovieLens-20M and
Netflix.

\end{document}